\documentclass[9pt,twocolumn,twoside]{article}

\usepackage[margin=1.8cm,columnsep=0.6cm]{geometry}
\usepackage{amsmath,amssymb,bm,cancel}
\usepackage{graphicx}
\usepackage{xcolor}
\usepackage{booktabs}
\usepackage{hyperref}
\usepackage[
  backend=biber,
  style=ieee,
  maxbibnames=3,
  minbibnames=3,
  url=false,
  doi=false,
  isbn=false,
  eprint=false
]{biblatex}

\AtEveryBibitem{%
  \clearlist{language}%
  \clearfield{langid}%
  \clearfield{origlanguage}%
  \clearfield{number}%
  \clearfield{issn}%
  \clearfield{doi}%
  \clearfield{url}%
  \clearfield{urldate}%
  \clearfield{isbn}%
  \clearfield{eprint}%
}
\usepackage{microtype}
\usepackage{caption}
\usepackage{titlesec}
\usepackage{parskip}
\usepackage{enumitem}
\usepackage{stfloats}
\usepackage{makecell}

\titleformat{\section}{\normalsize\bfseries\uppercase}{}{0em}{}
\titleformat{\subsection}{\normalsize\itshape}{}{0em}{}
\titlespacing*{\section}{0pt}{6pt}{2pt}
\titlespacing*{\subsection}{0pt}{4pt}{1pt}

\newcommand{\bsigma}{\bm{\sigma}}
\newcommand{\bgdot}{\dot{\bm{\gamma}}}
\newcommand{\bF}{\bm{F}}

\newcommand{\bI}{\bm{I}}
\newcommand{\bT}[1]{\bm{T}_{#1}}
\newcommand{\Wi}{\mathrm{Wi}}
\newcommand{\dd}{\mathrm{d}}

\begin{document}

\twocolumn[
\begin{center}
{\large\bfseries
Stable and Interpretable Multi-Mode Rheological
Universal Differential Equations (mmRUDEs) for
Data-Driven Constitutive Modeling}\par\vspace{6pt}
{\normalsize
Mohua Das\textsuperscript{1,\textdagger}, Nicholas King \textsuperscript{2,\textdagger}, Navid Azizan $^{3}$, Gareth H. McKinley \textsuperscript{1,*}}\par\vspace{3pt}

\textit{$^1$ Hatsopoulos Microfluids Laboratory, Department of Mechanical Engineering, Massachusetts Institute of Technology, Cambridge, MA 02139}\par\vspace{3pt}
\textit{$^2$ Department of Chemical Engineering, Massachusetts Institute of Technology, Cambridge, MA 02142}\par\vspace{3pt}
\textit{$^3$ Department of Mechanical Engineering;
Institute for Data, Systems, \& Society (IDSS),
and  Laboratory for Information \& Decision Systems (LIDS),
Massachusetts Institute of Technology, Cambridge, MA 02139}\par\vspace{3pt}

\end{center}
\vspace{16pt}
]

\begingroup
\renewcommand{\thefootnote}{\fnsymbol{footnote}}
\footnotetext[2]{M.D. and N.K. contributed equally to this work.}
\endgroup

\begingroup
\renewcommand{\thefootnote}{\fnsymbol{footnote}}
\footnotetext[1]{To whom correspondence may be addressed: G.H.M \\ 
E-mail:~\textcolor{blue}{ gareth@mit.edu} \\}

\endgroup


\section*{Abstract}
\small
The Rheological Universal Differential Equation (RUDE) framework embeds neural networks within a frame-indifferent tensorial constitutive backbone and so enables data-driven discovery of complex material rheological behavior. However, the flexibility that makes the RUDE framework attractive also makes it difficult to deploy: the learned neural network correction can make the constitutive equation numerically stiff during training, distribute corrections across non-unique combinations of tensor-basis terms, and extrapolate poorly with increasing nonlinearity, especially for industrially relevant materials with broad relaxation spectra. We develop a stability- and reliability-focused multi-mode RUDE framework that addresses three key challenges: thermodynamic admissibility, out-of-distribution prediction, and interpretability. Logarithmic compression of the input invariants and a split-network architecture improve numerical conditioning during training, while a differentiable projection layer enforces the Clausius--Duhem inequality at every time step. We illustrate the framework with both synthetic data (a five-mode nonlinear Giesekus model) and experimental data on an entangled silicone polymer melt. Both models were trained on large-amplitude oscillatory shear (LAOS); they remain stable and accurate under unseen flow conditions, including steady shear, transient stress growth, and extensional flows, for which the unconstrained models may diverge numerically. To aid in interpretation of the contributions of each learned correction to the rheological behavior, we introduce the \textit{rheome}, a compact way of representing the dominant weighted tensor-basis contributions to the overall rheological behavior. These improvements to the RUDE framework lead to stable and interpretable trained models, paving the way for implementation in computational fluid dynamics simulations and for guiding the rational design of soft material processing operations.

\normalsize

\vspace{4pt}\hrule\vspace{6pt}
{\small\textbf{Keywords:} constitutive equation discovery
$|$ universal differential equation $|$ tensor-basis neural networks $|$ thermodynamic admissibility $|$ polymer rheology $|$ Weissenberg number
$|$ hard-constrained neural networks}

\vspace{4pt}\hrule\vspace{6pt}

\section*{Significance Statement}
\small
Machine-learning techniques have recently been utilized to discover constitutive equations for complex fluids and soft solids from rheometric data. With limited shear data, the learned model may not be unique: many nonlinear functional forms fit the observations equally well, yet predict markedly different behavior under unseen conditions. We address this challenge through an integrated training pipeline designed to produce stable, interpretable, rheologically consistent, and thermodynamically admissible digital fluid twins. The framework improves numerical conditioning, stabilizes training and extrapolation across decades of flow strength, and guides the optimizer toward a frame-indifferent constitutive equation consistent with the ground truth.

\begin{center}
    \textit{The proposed framework is designed to keep the learned constitutive equation accurate where data are available, stable where they are not, and interpretable throughout.}
\end{center}

\normalsize

\section*{Introduction}

Constitutive equations relate the tensorial state of stress within a material to its deformation history. Their accuracy determines the reliability of simulations of polymer processing operations \cite{larson_modeling_2015} or flows of complex fluids \cite{spagnolie_complex_2015,barrat_soft_2024}, which in turn guide the rational design of these ``soft materials'' \cite{ewoldt_designing_2022}. Classical models for polymeric liquids such as the Giesekus \cite{giesekus_simple_1982}, Phan-Thien--Tanner (PTT) \cite{phan-thien_new_1977}, or the Finitely Extensible Nonlinear Elastic (FENE) \cite{bird_polymer_1980} family of models each assume specific microstructural mechanisms for the stretch and relaxation of polymer chains, although the choice of optimal material model must ultimately be guided by experimental data. However, high-fidelity rheological data is sparse and expensive to collect. A typical dataset collected using a commercial rheometer covers only a few flow protocols over a limited window of strains and strain rates. Furthermore, no single model thus far is able to capture the behavior of a complex fluid in all possible flows \cite{khan_comparison_1987}. The gap between what classical rheological models can represent and what limited experimental data can constrain motivates a different, data-driven approach to the constitutive modeling of complex fluids. 

The rapid improvement in scientific machine learning tools over the last decade \cite{faroughi_physics-guided_2024,mangal_data-driven_2025} has led to several efforts to apply them in the discovery of appropriate constitutive equations for complex fluids \cite{mangal_data-driven_2025}. Broadly, these models have flexibility in their parametric representation that span the range from fixed functional form models to symbolic library-based selection \cite{brunton_discovering_2016}, to structured non-parametric models such as neural Ordinary Differential Equations \cite{chen_neural_2018}, and finally to black box neural networks. For example, John \textit{et al.} \cite{john_machine_2024} applied the random forest algorithm to select the best parameters for fitting to various known rheological models characterized by only one or two nonlinear constants. On the other hand, Rheology-Informed Neural Networks (RhINNs), proposed by Jamali and co-workers \cite{mahmoudabadbozchelou_unbiased_2024}, are an adaptation of Physics-Informed Neural Networks (PINNs) \cite{raissi_physics-informed_2019} that learn constitutive equations using a black-box neural network. Recurrent neural networks based on the deformation history of the material have also been explored \cite{jin_data-driven_2023}. A good balance between discovery and interpretability for these constitutive models lies somewhere between these two approaches, embedding physical principles yet remaining sufficiently expressive to describe observed flow phenomena. 
Several complementary strategies have emerged for data-driven rheological modeling and constitutive discovery. Rheological Universal Differential Equations (RUDEs) embed learned constitutive corrections
within a tensorial differential equation framework \cite{lennon_scientific_2023}. A separate class of approaches either use sparse regression to identify parsimonious constitutive relationships \cite{shanbhag_sparse_2024,shanbhag_isolating_2026} or autoencoder-based reduced-order modeling. This class includes Rheo-SINDy \cite{sato_rheo-sindy_2025}, which applies the Sparse Identification of Nonlinear Dynamics (SINDy) algorithm \cite{brunton_discovering_2016} and the SIMPLE method \cite{young_scattering-informed_2023}, which applies autoencoders to identify invariant manifolds in high-dimensional dynamical systems.

For complex fluids, an important principle that constitutive equations must obey is the concept of material objectivity. This requires that the constitutive law must be unchanged under arbitrary time-dependent rigid-body rotations of the laboratory frame. This principle was formalized by Oldroyd in a seminal paper \cite{oldroyd_formulation_1950}, and has been explicitly incorporated as part of the machine learning architectures of the promising approaches detailed above. One way of ensuring that constraints such as material objectivity are obeyed is to use the Universal Differential Equation (UDE) framework, introduced by Rackauckas \cite{rackauckas_universal_2021}. UDEs retain the known physics, such as the symmetry and structure of the original ODE, and augment it with a neural network trained jointly with a numerical solver for the differential equation through automatic differentiation. This ensures that the learned correction is consistent with the dynamics, rather than fitting an isolated input-output map. In particular, RUDE \cite{lennon_scientific_2023} retains the core feature of the frame indifferent Oldroyd-B equation as the backbone, with tensorial nonlinear terms learned by a neural network, hence ensuring that material objectivity is always satisfied. The learned nonlinear corrections are expanded in a finite set of tensor bases using a representation theorem first proposed by Spencer and Rivlin \cite{spencer_theory_1958}. In its original form, a RUDE is a single (tensorial) differential equation, representing a viscoelastic fluid with a linear rheological response that can be adequately described by a single characteristic relaxation time. Using curriculum learning \cite{bengio_curriculum_2009} on single-mode synthetic data generated from the Giesekus model \cite{giesekus_simple_1982}, Lennon \textit{et al.} \cite{lennon_scientific_2023} showed that the network can recover known constitutive closures from only a few observations of a single component of the stress tensor in large amplitude oscillatory shear (LAOS) flow. They also showed that a trained RUDE can reproduce the data for a metal cross-linked hydrogel that has a narrow relaxation spectrum well described by a single relaxation mode. Subsequently, Rodrigues \textit{et al.} \cite{rodrigues_finding_2025} applied a similar strategy to identify a wider range of viscoelastic constitutive equations, further demonstrating the potential of this approach. Both Lennon \textit{et al.} \cite{lennon_scientific_2023} and Cummings \textit{et al.} \cite{cummings_harnessing_2026} have shown how single-mode learned RUDEs can be directly used for computational fluid dynamics simulations, with good agreement.

However, material objectivity alone does not guarantee stable or physically realistic extrapolation beyond the training regime. In shear flow, several of the basis tensors become nearly co-linear, implying that many different parameter combinations learned by the neural network can reproduce the training equally well, yet may produce wildly different and potentially divergent predictions under unseen flow conditions. This is a critical shortcoming when the goal is a broadly deployable constitutive model, or ``digital twin'' \cite{mahmoudabadbozchelou_unbiased_2024} capable of accurate predictions of non-Newtonian flow behavior in unseen flow configurations. The challenge is compounded when these models are applied to many commercial materials, which require multi-mode descriptions to accurately describe their relaxation dynamics \cite{baumgaertel_relaxation_1990}. The relaxation spectra of real complex fluids typically span orders of magnitude in relaxation time scales, and in our experience, gradient updates for scientific machine learning approaches frequently become dominated by the modes with the largest stress contribution, effectively losing the ability to learn nonlinear corrections across all relaxation time scales simultaneously. Another challenge is that the numerical blow-up of higher-order corrective terms can occur because certain basis tensors in the underlying tensorial representation theorem scale as the square of the stress tensor or higher. Hence, learned coefficients that are ostensibly harmless at low strain rates can trigger finite-time blow-up when the model is utilized in stronger flows either during later stages of training or at inference. 

Our contribution in this work is twofold. First, we reformulate the single-mode RUDE framework of Lennon \textit{et al.} \cite{lennon_scientific_2023} to incorporate multiple modes, which allows representation of the viscoelastic spectrum of real materials more accurately. Second, we add three ingredients that keep training stable, accurate, and interpretable. We replace the inputs to the neural network with logarithmically scaled values, effectively bringing all inputs onto a common scale so that gradient updates reflect contributions from all modes equally, rather than being dominated by a single term. Furthermore, we train two neural networks simultaneously rather than one to produce the scalar coefficients multiplying the basis tensor. This promotes a simplicity bias towards lower-order corrections and increases training stability. Finally, we append to the output of the neural network a differentiable projection layer that enforces the Clausius-Duhem inequality from non-equilibrium thermodynamics \cite{ottinger_beyond_2005,anand_continuum_2020}. These new contributions together convert the flexible but fragile RUDE architecture into a framework that generates nonlinear frame-indifferent constitutive equations for soft materials that are stable, accurate, and interpretable.

\section*{Framework}

\subsection*{Multi-mode Rheological Universal Differential Equations (mmRUDEs)}

We extend the single-mode RUDE framework of Lennon \textit{et al.}~\cite{lennon_scientific_2023} to $M$ upper convected Maxwell modes. Each mode $m$ is characterized by a relaxation time $\tau^{(m)}$, an elastic modulus $G^{(m)}$ (and a corresponding viscosity $\eta^{(m)}=G^{(m)}\tau^{(m)}$), and contributes a mode-specific tensorial stress $\bsigma^{(m)}$. All modes are subject to the same strain rate history, $\bgdot(t')$, and use a mode-specific neural correction with shared network parameters (weights and biases) $\bm{\theta}$. The total stress is 

\begin{align}
\bsigma=\sum_{m=1}^{M}\bsigma^{(m)}+\bsigma_{\text{solvent}},
\qquad
\bsigma_{\text{solvent}}=\eta_s\bgdot .
\end{align}

\begin{figure*}[!t]
\centering
\includegraphics[width=\textwidth]{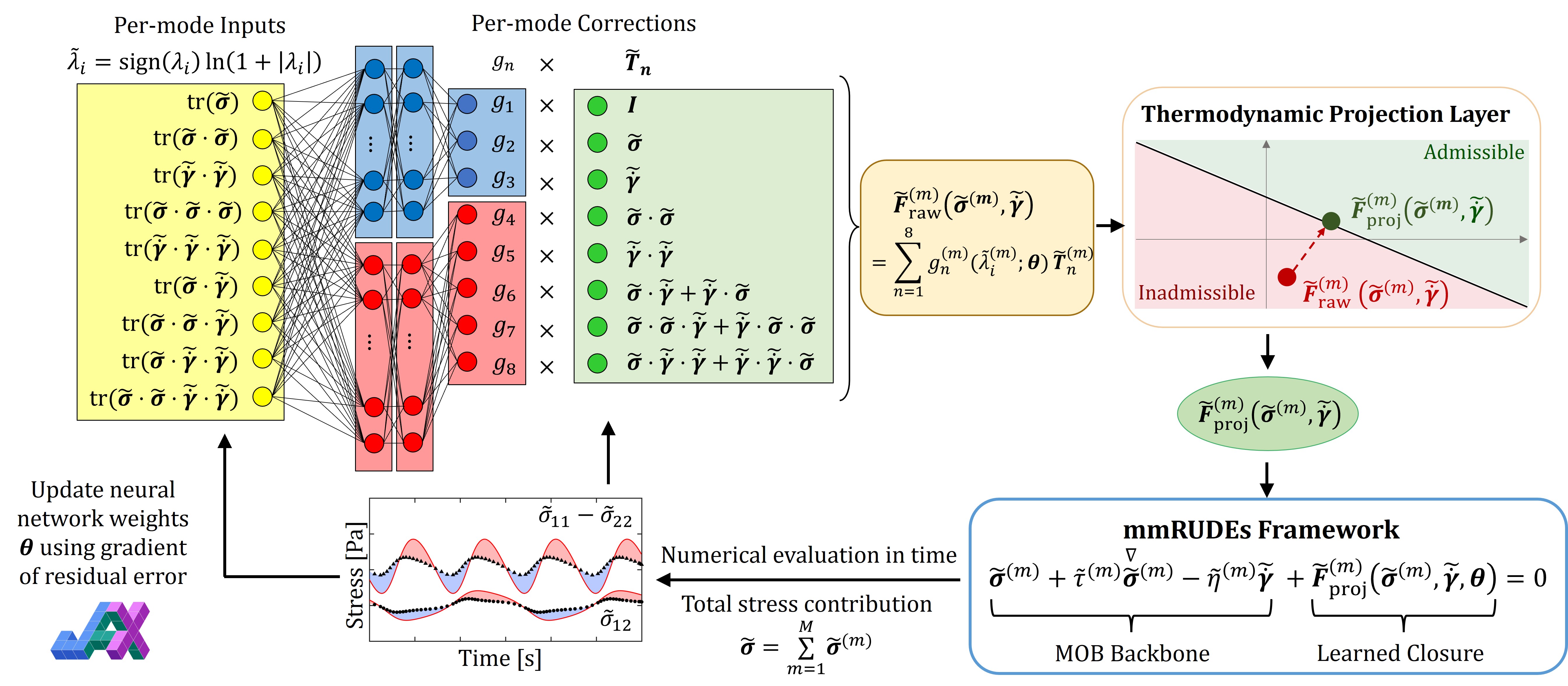}
\caption{\textbf{Rheologically consistent and thermodynamically admissible multi-mode Rheological Universal Differential Equations (mmRUDEs) framework.} For each relaxation mode $m$, nine scalar invariants of the normalized stress $\tilde{\bsigma}^{(m)}$ and strain rate $\tilde{\bgdot}$ are assembled (yellow vector) and compressed by the log-modulus transform $\tilde{\lambda}_i = \mathrm{sign}(\lambda_i)\ln(1+|\lambda_i|)$ [where $\lambda_i = \mathrm{tr}(\tilde{\bsigma}^{(m)}), \mathrm{tr}(\tilde{\bsigma}^{(m)}\cdot \tilde{\bsigma}^{(m)}), \cdots$], which maps invariants spanning many decades onto an $\mathcal{O}(10)$ range for the datasets considered. The compressed invariants are processed by two parallel sub-networks that share the same inputs: sub-network $\mathbb{NN}_{1}$ (blue; no bias) outputs the scalar functions $g_1$ to $g_3$, multiplying $\{\bI,\ \tilde{\bsigma}^{(m)},\ \tilde{\bgdot}\}$ and gives isotropic and linear corrections, while sub-network 2 i.e., $\mathbb{NN}_{2}$ (red; with bias) outputs $g_4$ to $g_8$, multiplying the higher order bases $\bT{4}$ to $\bT{8}$. The coefficients and tensor basis combine into the raw correction $\bF_{\mathrm{raw}}^{(m)} (\tilde{\bsigma}^{(m)}, \tilde{\bgdot}) $ for each mode $m$. The \emph{thermodynamic projection layer} then enforces the Clausius--Duhem dissipation inequality, projecting the learned $\bF$ term onto the admissible half-space if it violates the constraint. 
Each mode advances its stress with a Maxwell Oldroyd-B (MOB) backbone enhanced by the admissible correction, $ \bF_{\mathrm{proj}}^{(m)}(\tilde{\bsigma}^{(m)}, \tilde{\bgdot})$, and the modal stresses are summed to give the total stress $\tilde{\bsigma} = \sum_{m=1}^{M}\tilde{\bsigma}^{(m)}$. Integrating in time gives the predicted shear stress $\sigma_{12}$ and first normal-stress difference $\sigma_{11}-\sigma_{22}$ (black markers represent the ground-truth; shaded band shows the learned neural network correction relative to the MOB backbone), whose residual against the data is back-propagated through the differentiable solver (\texttt{JAX} autodiff) to update the parameters $\bm{\theta}$. The projection layer guarantees thermodynamic admissibility at every solver step, in both training and deployment.}

\label{fig:architecture}
\end{figure*}

Here, the solvent contribution is represented as a Newtonian dashpot with viscosity $\eta_s$. Equivalently, it may be viewed as the limiting response of a mode with an infinitesimally small relaxation time, $\tau^{(m)}\rightarrow 0$. Since $\bgdot$ is objective, the Newtonian stress contribution $\eta_s\bgdot$ is also frame-indifferent. The evolution equation describing the constitutive response of each remaining modal contribution to the stress is given by

\begin{equation}
\begin{split}
  \underbrace{
    \bsigma^{(m)}
    + \tau^{(m)}
      \overset{\nabla}{\bsigma}{}^{(m)}
    - \eta^{(m)}\,\bgdot
  }_{\text{MOB backbone}}
  +
  \underbrace{
    \bF^{(m)}
    \bigl(\bsigma^{(m)},\bgdot;\bm{\theta}\bigr)
  }_{\text{Learned closure}}
  &= \bm{0},
\end{split}
\label{eq:mob}
\end{equation}
where $\overset{\nabla}{(\cdot)}$ denotes the upper-convected derivative \cite{stone_note_2023}. 
Setting
$(\bF^{(m)}=\bm{0})$ recovers a classical multi-mode quasilinear model corresponding to a ladder model with $M$ frame-indifferent upper convected Maxwell modes. For compactness, we refer to this as the Maxwell––Oldroyd-B (MOB) framework \cite{shaqfeh_oldroyd-b_2021}. The neural network ($\mathbb{NN}$) correction (i.e., the learned closure) shown in Eq.~(\ref{eq:mob}) extends the MOB constitutive equation by introducing mode-specific constitutive correction terms. We express the correction for each mode as a tensor-basis expansion,
\begin{equation}
\bF^{(m)} = \sum_{i=1}^{8}
g_i^{(m)}\left({\bm{\lambda}}^{(m)};\bm{\theta}\right)\bT{i}^{(m)},
\label{eq:tbnn}
\end{equation}
where the eight scalar functions $g_i$ depend on the $\mathbb{NN}$ parameters $\bm{\theta}$ and nine scalar invariants, denoted $\bm{\lambda}^{(m)}$ of $(\bsigma^{(m)},\bgdot)$:
$\mathrm{tr}(\bsigma^{(m)})$,
$\mathrm{tr}(\bsigma^{(m)} \cdot \bsigma^{(m)})$,
$\mathrm{tr}(\bgdot \cdot \bgdot)$,
$\mathrm{tr}(\bsigma^{(m)} \cdot \bsigma^{(m)} \cdot \bsigma^{(m)})$,
$\mathrm{tr}(\bgdot \cdot \bgdot \cdot \bgdot)$, and four mixed traces [see representation in Fig.~\ref{fig:architecture}].

Each tensor basis $\bT{i}$ is constructed from independent symmetric products of the two $(3\times3)$ symmetric tensor arguments, $\bsigma$ and $\bgdot$. By the Cayley--Hamilton theorem, only powers $(p,q\in\{0,1,2\})$ need be considered. Because the stress $\bsigma$ and deformation rate $\bgdot$ tensors do not generally commute, mixed products are symmetrized as
$[
\bsigma^{p}\cdot\bgdot^{q}
+
\bgdot^{q}\cdot\bsigma^{p}
]$, whereas pure powers of a single tensor are already symmetric. The bases are therefore written
\begin{align}
  \{\bT{i}\}_{i=1}^8 =
  \{&\bI,~\ \bsigma,~\ \bgdot,~\
    \bsigma\!\cdot\!\bsigma,~\ \bgdot\!\cdot\!\bgdot,~(\bsigma\!\cdot\!\bgdot+\bgdot\!\cdot\!\bsigma),\nonumber\\
    &(\bsigma\!\cdot\!\bsigma\!\cdot\!\bgdot +\bgdot\!\cdot\!\bsigma\!\cdot\!\bsigma),~~(\bsigma\!\cdot\!\bgdot\!\cdot\!\bgdot
+\bgdot\!\cdot\!\bgdot\!\cdot\!\bsigma)\}.
  \label{eq:tensor_basis}
\end{align}

The highest-order basis tensor previously considered by Lennon \textit{et al.}\cite{lennon_scientific_2023}, corresponding to the $(p,q)=(2,2)$ entry,
$[
\bsigma\cdot\bsigma\cdot\bgdot\cdot\bgdot
+\bgdot\cdot\bgdot\cdot\bsigma\cdot\bsigma
]$
was recently shown to be unnecessary, as this term can be expressed in terms of the other eight basis tensors \cite{kamrin_clarifying_2025}. It is therefore omitted, leaving the remaining eight basis tensors, denoted by $(\bT{1},\ldots,\bT{8})$, as the tensor bases used in Eq.~\ref{eq:tbnn}.
 
These basis tensors $\bT{i}$ and the associated $g_i$ coefficients are sufficiently expressive to represent the nonlinear constitutive terms appearing in a wide range of standard constitutive models. Among phenomenological models, the Giesekus \cite{giesekus_simple_1982} nonlinearity is represented by $\bT{4}$, whereas the Johnson--Segalman model \cite{johnson_model_1977} involves $\bT{6}$. The linear and exponential Phan--Thien--Tanner (PTT) models \cite{phan-thien_new_1977} involve contributions associated with $\bT{2}$ and $\bT{6}$. The microstructurally motivated Rolie--Poly model \cite{likhtman_simple_2003} draws on $\bT{1}$ and $\bT{2}$.
Full constitutive equations for each of these models are provided in SI~Appendix~F. In summary, the set of proposed tensor bases can completely represent many of the physically important nonlinear terms that typically appear in an unknown material constitutive equation.

\subsection*{Stable training}

Training a multi-mode RUDEs framework is numerically challenging because real materials are characterized by mode-specific parameters $(G^{(m)}, \tau^{(m)})$ that span several decades in magnitude \cite{bird_dynamics_1987}. The stress generated over an oscillatory deformation cycle, and the values of the input invariants $(\lambda_{1},\cdots,\lambda_{9})$ also typically range over many orders of magnitude. This creates two main
challenges: poorly scaled network inputs and rapidly growing higher-order basis tensor contributions, both of which stiffen the ordinary differential equations (ODEs) produced by the RUDE framework (Eq.~(\ref{eq:mob}) and destabilize training. In addition,
unconstrained optimization of the loss function (as described in the subsequent subsection) may distort the underlying MOB backbone contained in Eq.~(\ref{eq:mob}). We address these issues using the following approaches: (i) a characteristic-scale nondimensionalization of all
dynamic variables, (ii) a sign-preserving logarithmic compression of the network inputs, and (iii) a split-network architecture.

All dynamic variables are first nondimensionalized by a characteristic modulus and timescale $(G_c,\tau_c)$, i.e., by defining a dimensionless time $\tilde t=t/\tau_c$, stress $\tilde{\bsigma}=\bsigma/G_c$, and strain rate
$\tilde{\bgdot}=\tau_c\,\bgdot$. The characteristic scales $(G_c,\tau_c)$ used for nondimensionalization
should be regarded as conditioning hyperparameters rather than uniquely
defined material properties.
A rheologically motivated default choice for this scaling is the viscosity-weighted (second-moment) relaxation time of the underlying discrete relaxation spectrum and a modulus defined as the zero-shear viscosity divided by this time scale, (which is equivalent to the inverse of the steady-state shear compliance $J_e^0$ from linear viscoelasticity \cite{white_dynamics_1964}). 

\begin{equation}
  \tau_c \equiv \bar{\tau}_\eta
  = \frac{\sum_{m=1}^{M} \eta^{(m)}\,\tau^{(m)}}
         {\sum_{m=1}^{M} \eta^{(m)}},
\quad
G_{\mathrm c}\equiv 1/J_e^{0}=\frac{\sum_{m=1}^M\eta^{(m)}}{\tau_{\mathrm c}};
\label{Viscosity weighted mean}
\end{equation}
However, other choices for the scaling
may be used when they improve the numerical conditioning of the multi-mode system. The selected values must remain fixed during a given
training and deployment calculation.

For an arbitrary flow with invariant magnitude $\lambda_3 = \text{tr} (\pmb{\dot{\gamma}} \cdot \pmb{\dot{\gamma}})$, a scalar strain rate can be defined generally as $\dot{\gamma} = \sqrt{\lambda_3/2}$. To measure flow strength, we define the Weissenberg number \cite{poole_deborah_2012} as $\mathrm{Wi} = \bar{\tau}_\eta \dot{\gamma}$, and for oscillatory shear flows at an angular frequency $\omega$, the Deborah number \cite{poole_deborah_2012, reiner_deborah_1964}  $\mathrm{De} = \bar{\tau}_\eta \omega$ is a measure of flow unsteadiness.

Even after nondimensionalization, the nine invariants $\lambda_i$ include traces of tensor products up to fourth order [e.g.,\ $\mathrm{tr}(\tilde\bsigma\cdot\tilde\bsigma\cdot\tilde\bgdot\cdot\tilde\bgdot)$], so the values of the higher-order basis tensors can potentially span over many decades in magnitude. Passed directly to the neural network, such inputs can saturate the activation functions and yield ill-conditioned gradients. We therefore compress each invariant with the sign-preserving log-modulus transform
\begin{equation}
  \tilde{\lambda}_i = \mathrm{sign}(\lambda_i)\,\ln\!\left(1+|\lambda_i|\right),
  \label{eq:logmod}
\end{equation}
which maps the inputs onto an $\mathcal{O}(10)$ compressed range while preserving their sign and monotonic ordering, so the network sees well-conditioned inputs over the range of stress or flow amplitude considered here. We note here that an alternative possible approach would be to reformulate the entire tensorial constitutive equation using the log-conformation tensor form \cite{fattal_constitutive_2004,fattal_time-dependent_2005}, but this represents a significant revision of the basic RUDE architecture, and we do not pursue it further in this work.

The identity basis $\bT{1}=\bI$ requires special conditioning at equilibrium. In the rest state $(\bgdot=0)$, we expect that the extra stress for any complex fluid is $\bsigma=\bm{0}$, consequently all nine invariants vanish and hence $\tilde{\lambda}_i=\bm{0}$, whereas $\bT{1}$ remains nonzero. A nonzero value of $g_1(\bm{0})$, for example due to a finite neural network bias, would therefore introduce an unphysical isotropic contribution to the extra stress tensor. To enforce a vanishing correction at rest, the coefficients of the basis tensors are generated by two subnetworks:
\begin{align}
  &\mathbb{NN}_{1}:\ (g_1,g_2,g_3)
    = \bm{W}_3^{(1)}\,\phi\!\bigl(\bm{W}_2^{(1)}\,
       \phi(\bm{W}_1^{(1)}\tilde{\bm{\lambda}})\bigr), \\
  &\mathbb{NN}_{2}:\ (g_4 \cdots g_8)
    = \bm{W}_3^{(2)}\,\phi\!\bigl(\bm{W}_2^{(2)}\,
       \phi(\bm{W}_1^{(2)}\tilde{\bm{\lambda}}+\bm{b}_1)+\bm{b}_2\bigr)+\bm{b}_3,
\end{align}
where $\phi$ is a smooth hidden activation with $\phi(0)=0$ (e.g. \texttt{gelu}, \texttt{tanh}, or \texttt{softsign} functions \cite{goodfellow_deep_2016}) and $\bm{W}_j^{(k)}$ and $\bm{b}_j$ are the weight matrices and bias vectors of the neural network $\mathbb{NN}_{k}$ of depth $j$.. The subnetwork $\mathbb{NN}_{1}$ carries \emph{no biases}, so at $\tilde{\bm{\lambda}}=\bm{0}$ each layer returns $\phi(\bm{0})=\bm{0}$ and hence $g_1=g_2=g_3=0$ exactly---the correction vanishes at rest by construction. The subnetwork $\mathbb{NN}_{2}$ generates the coefficients
$g_4$ to $g_8$, and the underlying higher-order basis tensors  $\bT{4}$ to $\bT{8}$ already vanish at rest. Although strictly only $g_1$ requires the bias-free constraint, we group all first-order linear tensorial bases (including $g_2\bT{2},g_3\bT{3}$) together in $\mathbb{NN}_1$ as a conservative choice.

This split helps training in three ways. First, the lower-order tensor basis contributions $g_1\bT{1}$ to $g_3\bT{3}$ and higher-order tensor basis contributions $g_4\bT{4}$ to $g_8\bT{8}$ have decoupled parameters, thereby reducing direct gradient competition between the fast-growing high-order terms and the smaller linear corrections. Second, under staged training, low-amplitude linear response data initially emphasize subnetwork $\mathbb{NN}_{1}$, while subnetwork $\mathbb{NN}_{2}$ stays near its initialization, and higher-amplitude data later populates $\mathbb{NN}_{2}$ while largely preserving the linear response already captured by $\mathbb{NN}_{1}$. Third, the high-order coefficients can be regularized more aggressively without suppressing the lower-order ones. Finally, the output weight matrices $\bm{W}_3^{(1)},\bm{W}_3^{(2)}$ are initialized at or near zero, so $\bF\approx\bm{0}$ at the start of training and the initial material response is always consistent with the quasi-linear material response predicted by the MOB backbone and the memory integral expansion for complex materials \cite{bird_dynamics_1987}; the correction appears only as these weights depart from zero with increasing non-linearity.

\subsection*{Thermodynamic projection: reliable training and deployment}
\label{sec:projection}
Although the Maxwell Oldroyd-B backbone ensures material objectivity, it does not guarantee that the learned constitutive correction will remain reliable or physically sensible when extrapolated beyond the training data. The stable training pipeline with log-compression and subnetworks keeps the optimization well-behaved, but does not decide which tensor bases contributions are selected. Importantly, neither of these improvements to the underlying RUDE framework ensures the thermodynamic admissibility, and hence reliability, of the learned correction, which is the most important and non-negotiable criterion for any data-driven constitutive modeling. 

High-fidelity training datasets primarily consist of data obtained in transient shear flows, which are the most accessible experimentally from a rheometer, and therefore probe only a subset of the full tensorial response. As a result, multiple combinations of the eight basis tensor contributions given in Eq.~\ref{eq:tensor_basis} may reproduce the observations equally well, with no intrinsic mechanism directing the optimizer toward thermodynamically admissible solutions. In particular, the optimizer may assign more weight to the higher-order basis $\bT{4}$ to $\bT{8}$, whose magnitudes grow more rapidly with increasing non-linearity. Such a model may fit low-amplitude data well, while becoming unstable as higher-Weissenberg-number datasets are introduced during the continual training procedure. More importantly, this learned model may generate unphysical stress contributions when evaluated using deformation rate histories outside the training distribution.

Reliability of the closure is therefore important, and this must hold both during continual training and throughout later deployment. There have been several recent successful efforts in machine learning to design hard-constrained neural networks to enforce physical or structural constraints on model predictions \cite{tang_learning_2025,goertzen_eco_2025}. Following the hard-constrained neural-network construction, HardNet, of Min and Azizan \cite{min_hardnet_2025}, we enforce these requirements as a \emph{hard} constraint rather than a \emph{soft} penalty: a projection layer maps every raw network output onto the thermodynamically admissible set before it enters the tensorial stress ODE. Because the admissible set depends on the current stress state, the constraint is input-dependent at each time step. The optimizer thus remains free to fit the data, but only with constitutive corrections that are thermodynamically admissible.

\textbf{Free energy and the dissipation constraint.}
For this work, we specifically model polymeric liquids, which have a well-established thermodynamic framework for analysis \cite{ottinger_beyond_2005}. For example, consider a dilute polymer solution modeled as a suspension of elastic dumbbells \cite{bird_polymer_2016}. The configuration of each dumbbell is described by its conformation tensor $\pmb{c}$, which is a symmetric positive-definite rank-2 tensor that reduces to $\bI$ at equilibrium. The Helmholtz free energy per unit volume of this fluid is hence \cite{nieto_simavilla_hammering_2025}:
\begin{equation}
  \rho\Psi = \frac{G}{2}\bigl[\,\Phi(\mathrm{tr}\,\pmb{c})
            - \ln\det\pmb{c} - 3\,\bigr],
  \label{eq:free_energy}
\end{equation}
where $\rho$ is the density, $G = nk_B T$ is the elastic modulus, which depends on the number density of dumbbells $n$, Boltzmann constant $k_B$, and the temperature $T$, and $\Phi(\text{tr}~\pmb{c})$ is an elastic (entropic) spring potential whose derivative defines the dimensionless spring function $f \equiv \dd\Phi/\dd(\mathrm{tr}\,\pmb{c})$. In the case of a Hookean spring, which covers both the Oldroyd-B and Giesekus models, the spring function $f=1$. To account for finite chain extensibility effects at high chain stretch, the FENE-P spring function $f = (b~{-}~3)/(b - \mathrm{tr}\,\pmb{c})$ \cite{bird_polymer_1980} can alternatively be chosen, where $b = \text{max}\{\text{tr} ~\pmb{c}\}$ is the finite extensibility of the chain.

Differentiating Eq.~\ref{eq:free_energy} in time, applying the stress--conformation relation $\bsigma = G(f\pmb{c} - \bI)$, and substituting the stress ODE [Eq.~\ref{eq:mob}] reduces the Clausius--Duhem inequality \cite{anand_continuum_2020}, expressed as $\rho\dot\Psi \leq \bsigma:\nabla\bm{v}$ (where $\nabla\bm{v}$ is the velocity gradient tensor), to a single algebraic condition on the learned correction (the full algebraic development is provided in SI~Appendix~A):
\begin{equation}
  (f\,\bI - \pmb{c}^{-1}):(\bsigma + \bF) \geq 0,
  \label{eq:constraint}
\end{equation}
with $\pmb{c} = (\pmb{I} + \pmb{\sigma}/G)/f$. For the remainder of this work, we only consider the neo-Hookean limit of $f = 1$. Thus, the thermodynamic projection adopted below assumes that \(f\) is constant so that \(\dot{f}=0\); by contrast, nonlinear spring relations of the form \(f=f(\mathrm{tr}\,\mathbf{c})\) would necessitate retaining the extra \(\dot{f}\,\mathbf{c}\) term in the rate of change, \(\dot{\boldsymbol{\sigma}}=G(f\dot{\mathbf{c}}+\dot{f}\,\mathbf{c})\).

\textbf{Minimum-norm projection.}
Eq.~\ref{eq:constraint} defines a feasible half-space in the six-dimensional space of symmetric $3\times3$ tensors. Its boundary is the hyperplane $(f\,\bI - \pmb{c}^{-1}):(\bsigma + \bF) = 0$. When the raw output $\bF_\mathrm{raw}$ lies within the admissible subspace, no correction is needed. When it lies outside, we project back onto the boundary to minimally satisfy the constraint:
\begin{equation}
 \bF_{\mathrm{proj}} = \bF_{\mathrm{raw}}
    + \frac{(f\,\bI - \pmb{c}^{-1})}{\|(f\,\bI - \pmb{c}^{-1})\|^2}\,
      \texttt{ReLU}~\!\bigl(-(f\,\bI - \pmb{c}^{-1}):(\bsigma + \bF_{\mathrm{raw}})\bigr).
  \label{eq:projection}
\end{equation}
The \texttt{ReLU} function ensures that there is a correction term only when the constraint is violated, and the prefactor $(f\,\bI - \pmb{c}^{-1})/\|(f\,\bI - \pmb{c}^{-1})\|^2$ moves $\bF_\mathrm{raw}$ along the constraint normal by the right amount to land on the boundary, resulting in the closest feasible $\bF$. 
Substituting Eq.~\ref{eq:projection} into the constraint gives exactly
$(f\,\bI - \pmb{c}^{-1}):(\bsigma + \bF_{\mathrm{proj}}) \geq 0$, guaranteeing that thermodynamic admissibility is observed at every time step. Fig.~\ref{fig:architecture} provides a schematic overview of the proposed framework, highlighting its stability, reliability, rheological admissibility, and thermodynamic consistency.

\subsection*{Training}
\label{sec:training}

The network weights, which are shared across modes, are trained to reproduce the reference stress trajectories. The observed quantities may include both the time-evolving shear stress $(\sigma_{12}(t))$ and the first normal-stress difference $(N_1(t)=\sigma_{11}(t)-\sigma_{22}(t))$, and, when available, the second normal-stress difference $(N_2(t)=\sigma_{22}(t)-\sigma_{33}(t))$. We minimize the normalized trajectory loss
\begin{align}
\begin{split}
  \mathcal{L}(\bm{\theta})
  &= \sum_{p}\!\left[
      \frac{\big\lVert \sigma_{12,p}^{\mathrm{obs}}-\sigma_{12,p}^{\mathrm{pred}}\big\rVert^{2}}
           {\big\lVert \sigma_{12,p}^{\mathrm{obs}}\big\rVert^{2}}
      + w_{N_1}\,
      \frac{\big\lVert N_{1,p}^{\mathrm{obs}}-N_{1,p}^{\mathrm{pred}}\big\rVert^{2}}
           {\big\lVert N_{1,p}^{\mathrm{obs}}\big\rVert^{2}}
    \right]
    \\
    &\quad+ \Lambda\,\lVert\bm{\theta}\rVert^{2},
  \label{eq:loss}
\end{split}
\end{align}
where $p$ indexes the training datasets; the superscripts `$\mathrm{obs}$' and `$\mathrm{pred}$' respectively denote the ground-truth and model-predicted trajectories; $\Lambda$ is the $L_2$ regularization coefficient; and $w_{N_1}$ weights the normal-stress residual relative to the shear stress. This is set to zero if training is performed only using the shear stress. Gradients are backpropagated through the differentiable ODE solver.

The protocols are introduced through a cumulative stagewise curriculum that gradually increases the deformation amplitude, i.e., increasing the nonlinearity of the constitutive response and hence difficulty \cite{bengio_curriculum_2009}. Since the correction starts at or near zero, training begins precisely with the MOB backbone. In the early stages, which involve low deformation amplitudes, only small, nearly linear adjustments are needed (and captured by subnetwork ${\mathbb{NN}_{1}}$); higher-order terms are activated only as more challenging protocols are introduced. As a result, the optimizer continuously develops the correction from the well-conditioned linear solution, rather than attempting to find a large nonlinear closure all at once. This process represents a well-posed progression from the MOB backbone of Eq.~(\ref{eq:mob}) to the higher-order fully nonlinear model, in which each stage refines the previous one instead of overwriting it. Neural network parameters are optimized with \texttt{Adam} \cite{kingma_adam_2015} with explicit \(L^2\) regularization, and each modal stress equation is advanced using an explicit adaptive Runge--Kutta scheme.

\begin{figure*}[!t]
\centering
\includegraphics[width=0.9\textwidth]{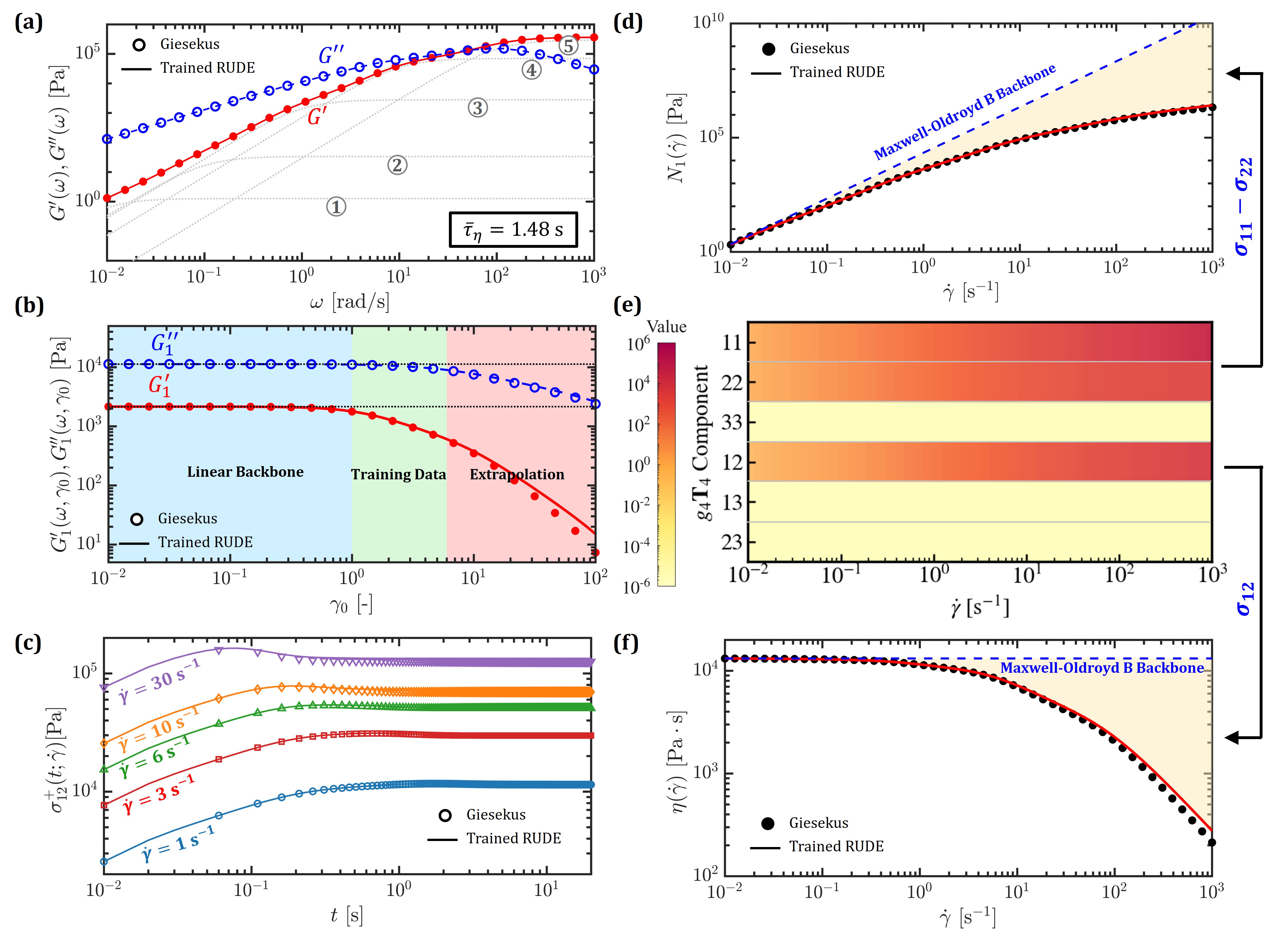}
\caption{\textbf{Trained mmRUDEs reproduce the results of the ground truth multi-mode Giesekus model across oscillatory and steady flows. } In all panels, symbols represent the reference 5-mode Giesekus model (with model parameters given in the text), and lines show the trained mmRUDEs prediction. (\textbf{a})~Comparison of the dynamic moduli $(G'(\omega), ~G''(\omega))$ predicted by the trained model and the ground truth. Light gray lines represent the contribution of each mode to the total $G'(\omega)$ response; (\textbf{b})~Large-amplitude oscillatory shear (LAOS) at $\omega=1$~rad/s (De = 1.48) as the strain amplitude ($\gamma_0$) is varied. The green regime is the training window ($\gamma_0=1,3,4,6$); the blue regime (low amplitude) shows that the learned correction vanishes in the linear limit; the red regime (up to $\gamma_0=100$, $\mathrm{Wi}\approx148$) shows accurate extrapolation for nonlinear flows. (\textbf{c})~Shear-stress growth predictions for startup of steady shear test agree closely with the ground truth throughout the transient evolution and approach to steady state at times $t/\bar\tau_\eta \gg 1$. (\textbf{d})~Steady first normal stress difference $N_1(\dot\gamma)$ and (\textbf{f})~shear viscosity $\eta(\dot\gamma)$ over five decades of steady shear rate ($10^{-2} \leq \dot\gamma \leq 10^3$ s$^{-1}$, $0.0148 \leq \mathrm{Wi} \leq 1480$); (\textbf{e})~Heatmap of the six individual components of the dominant learned correction in steady shear flow $g_4\tilde{\bT{4}}$ ($\bT{4}=\tilde{\bsigma}\cdot\tilde{\bsigma}$)  across the five modes; the correction grows smoothly from the MOB backbone as the shear rate is increased and the nonlinearity of the flow increases.}
\label{fig:training giesekus}
\end{figure*}

\begin{figure*}[!t]
\centering
\begin{minipage}[t]{\textwidth}
  \centering
  \includegraphics[width=0.9\textwidth]{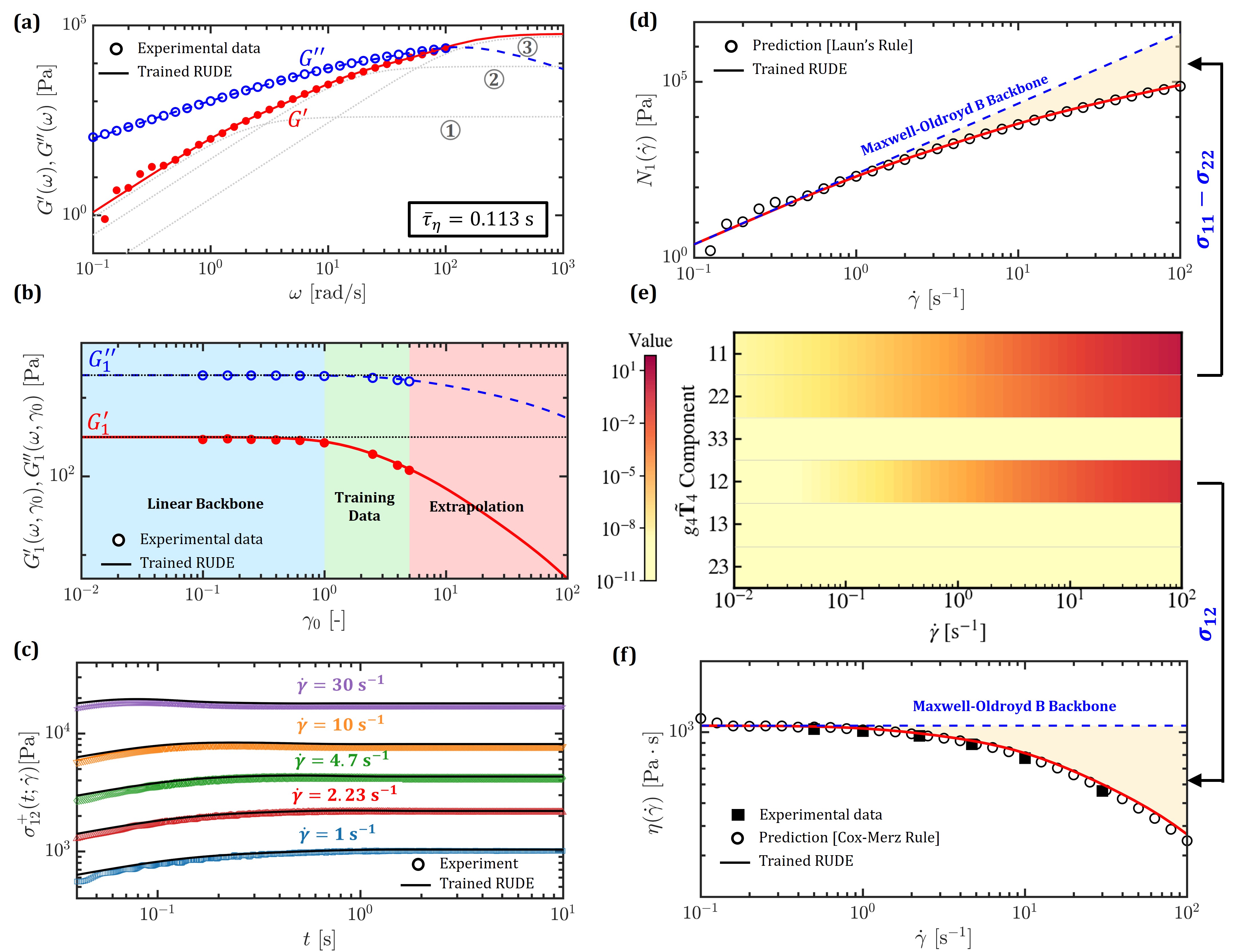}
\end{minipage}
\caption{\textbf{Trained mmRUDEs reproduce the results of the ground-truth experimental data of PDMS across oscillatory and steady flows.} In all panels, symbols represent the experimental data, lines show the trained mmRUDEs prediction. (\textbf{a})~Comparison of the linear viscoelastic spectra predicted by the trained model and the ground truth, showing the storage modulus $G'(\omega)$ and loss modulus $G''(\omega)$. Light gray lines show the contribution of each mode to the $G'(\omega)$ response; (\textbf{b})~Large-amplitude oscillatory shear (LAOS) at $\omega=2$~rad/s (De = 0.226) as the strain amplitude $\gamma_0$ is varied. The green regime is the training window ($\gamma_0=1,2.5,4,5$; $0.226 \leq \mathrm{Wi} \leq 1.13$); the blue regime (low amplitude) shows that the learned correction vanishes in the linear limit; the red regime (up to $\gamma_0=100$, $\mathrm{Wi} \approx22.6$) shows stable extrapolation for nonlinear flows. (\textbf{c})~Shear-stress growth predictions in the startup of steady shear flow agree closely with the ground truth throughout the transient evolution and the approach to steady state. (\textbf{d}) Evolution of the first normal stress difference $N_1(\dot\gamma)$ and (\textbf{f})~shear viscosity $\eta(\dot\gamma)$ in steady shear flow over three decades of steady shear rate ($10^{-1} \leq \dot\gamma \leq 10^{2}$ s$^{-1}$, $0.0113 \leq \mathrm{Wi} \leq 11.3$). Hollow symbols are the predictions using the Cox-Merz rule \cite{cox_correlation_1958} and Laun's rule \cite{laun_prediction_1986,das_launs_2024} respectively; (\textbf{e})~Heatmap of the dominant learned correction $g_4\tilde{\bT{4}}$ ($\bT{4}=\tilde{\bsigma}\cdot\tilde{\bsigma}$) across the three modes; the correction grows smoothly from the MOB backbone as the nonlinearity of the flow increases.
}
\label{fig:pdms}
\end{figure*}

\section*{Results}
\label{sec:results}

We demonstrate the capabilities of this new framework using two illustrative systems: synthetic data generated from a five-mode Giesekus model \cite{giesekus_simple_1982} (Fig.~\ref{fig:training giesekus}) and experimental data measured on a polydimethylsiloxane (PDMS) sample (Fig.~\ref{fig:pdms}). In each case, the model is tested on a sequence of progressively more challenging tasks---fitting the training data, extrapolating beyond the training strain-amplitude window, and generalizing to entirely unseen flow protocols. Full training and testing details are described in the following subsections.

\subsection*{Multi-mode Giesekus Model}
We first validate the proposed framework using synthetic data generated from a non-linear constitutive model. The ground truth is a five-mode non-linear Giesekus model \cite{giesekus_simple_1982} with mode-specific parameters $(G^{(m)},\tau^{(m)}) =$ (1.27~Pa, 100~s), (33.7~Pa, 10~s), (2800~Pa, 1~s), (70200~Pa, 0.1~s), and (295000~Pa, 0.01~s), and the same constant mobility parameter $\alpha=0.8$ for every mode. The relaxation-time spectrum spans four decades of time scale. We can characterize the spectrum using the viscosity-weighted mean relaxation time [Eq.~\ref{Viscosity weighted mean}], $\bar{\tau}_\eta \approx 1.48$~s. The corresponding linear viscoelastic response, represented by the storage modulus $G'(\omega)$ and loss modulus $G''(\omega)$ under small-amplitude oscillatory shear (SAOS), is shown in Fig.~\ref{fig:training giesekus}(a).

Here, the known part of the constitutive equation is the MOB backbone, while the network learns the Giesekus nonlinearity through the mode-specific corrections $\tilde{\mathbf F}^{(m)}$. When a Newtonian solvent is included, as in the Oldroyd--B framework, it contributes only a parallel contribution or viscous stress, $\tilde{\bsigma}_{\text{solvent}}=\tilde\eta_s\tilde{\bgdot}$, which is linear, objective, and dissipative. This solvent contribution is added algebraically to the total stress after solving the mode-wise constitutive equations and therefore lies outside both the learned correction and the thermodynamic projection. Thus, the training procedure is unchanged by the presence of a solvent: only the polymeric contribution receives learned, thermodynamically projected corrections.

The mmRUDEs framework is trained on large-amplitude oscillatory shear (LAOS) data at a single angular frequency $\omega=1$~rad/s and four progressively larger strain amplitudes $\gamma_0=1,3,4,6$, corresponding to Weissenberg numbers $\mathrm{Wi} =\bar{\tau}_\eta\,\gamma_0\,\omega \approx 1.5,\ 4.4,\ 5.9,\ \text{and}\ 8.9$ (the normalization parameters used during training are $G_c = 1200~ \text{Pa}$ and $\tau_c = 0.001 ~ \text{s}$). Both the oscillatory shear stress and first normal stress difference were used in this training, with weight $w_{N_1} = 1$ in the loss function. Within this trained window, the surrogate reproduces the Giesekus response almost exactly, as shown by the green shaded regime of Fig.~\ref{fig:training giesekus}(b).

To assess generalization, we evaluate the trained model on a strain-amplitude sweep extending both below and above the training range, as shown in Fig.~\ref{fig:training giesekus}(b). At low amplitudes (blue), the response approaches the linear viscoelastic (LVE) regime and the learned correction becomes negligible. The MOB backbone therefore remains undistorted, preserving accurate descriptions of the storage and loss moduli, $G'(\omega)$ and $G''(\omega)$, despite training on weakly nonlinear data. This consistency is also evident in Fig.~\ref{fig:training giesekus}(a), where the linear viscoelastic moduli predicted by the learned mmRUDE framework agree exactly with those of the reference multi-mode Giesekus model. 
At high amplitudes, we extend the model predictions to $\gamma_0=100$, corresponding to $\mathrm{Wi}\approx 148$. This amplitude is more than an order of magnitude larger than the most nonlinear training condition, $(\gamma_0=6$, corresponding to  $\mathrm{Wi}\approx8.9)$. Despite the limited training data provided, the learned surrogate model continues to reproduce the Giesekus response accurately in the strongly nonlinear regime. The constrained correction therefore extrapolates smoothly without exhibiting unbounded growth.

Another test of the reliability of the learned constitutive equation is prediction under flow conditions different from training. We evaluate the trained model's performance in both transient and steady shear flows. During the start-up of steady shear, the predicted evolution of the shear stress response follows the multi-mode Giesekus reference from the initial transient to the steady state, as shown in Fig.~\ref{fig:training giesekus}(c). To evaluate only the steady-state predictions, we sweep the shear rate over five decades, $\dot\gamma = 10^{-2}$ to $10^{3}$~s$^{-1}$ ($10^{-2} \lesssim\mathrm{Wi} =\bar{\tau}_\eta\dot{\gamma} \lesssim 10^3$), and compare the steady-state material functions with the reference data (symbols). The shear viscosity $\eta(\dot\gamma)$ (Fig.~\ref{fig:training giesekus}(f)) and the first normal-stress difference $N_1(\dot\gamma)$ (Fig.~\ref{fig:training giesekus}(d)) both agree with the multi-mode Giesekus reference across the entire range, capturing the shear-thinning of $\eta(\dot\gamma)$ and the growth of $N_1(\dot\gamma)$ that mark the nonlinear response. 

Beyond predictive accuracy, the learned correction is also physically interpretable. The exact Giesekus closure contains only one quadratic contribution,
\begin{equation}
  \bF^{(m)}_{\mathrm{Giesekus}}
  = \frac{\alpha}{G^{(m)}}\,\bsigma^{(m)}\!\cdot\!\bsigma^{(m)},
  \label{eq:giesekus}
\end{equation}
which corresponds to the tensor-basis term $(g_4\bT{4})$, with
$(\bT{4}=\bsigma\cdot\bsigma)$ and $g_4$ constant. A model that recovers the underlying constitutive equation from the trained dataset should therefore place most of the learned correction on this tensor basis. Figure~\ref{fig:training giesekus}(e) provides one graphical representation of this behavior using a heat map. The only non-zero components are the 11, 22, and 12 contribution in the dimensionless stress contribution $(g_4\tilde{\bT{4}})$, and the values increase smoothly from the MOB backbone as the imposed deformation becomes more nonlinear. A consistent pattern is exhibited across all five modes, while the remaining basis contributions stay small (less than 0.1\% of $g_4\tilde{\bT{4}}$). A detailed discussion and more sophisticated graphical representation of the individual tensor-basis contributions is provided in section: \textbf{Interpretability of mmRUDEs}.

Further validation examples of mmRUDEs trained on synthetic data are presented in the appendices. An ablation study that isolates the contribution of the projection layer to mmRUDE stability is provided in Appendix B. This ablation study confirms that models trained with the thermodynamic projection remain stable during extrapolation, whereas models trained without it become unstable and diverge. We further validate the framework on a multi-mode Giesekus fluid with mode-specific mobility parameters $\alpha^{(m)}$ (Appendix D), and also using a multi-mode exponential Phan–Thien–Tanner (ePTT) fluid (Appendix E). The ePTT case reveals an identifiability limitation inherent to simple shear. In simple shear, the $\bT{4}$ term $\bsigma\cdot\bsigma$ can also be written in terms of $\bI$ and $\bsigma$, so the $g_2\bT{2}$ and $g_4\bT{4}$ contributions become co-linear in simple shear and cannot be separated using analysis of shear stress and $N_1$ data alone. The most robust resolution would either be to include data for the second normal stress difference $N_2$ in the training datasets, although this is very difficult to access experimentally \cite{maklad_review_2021}, or to use different flow types (e.g., an extensional flow). Therefore, unique recovery of the constitutive equation may require more extensive datasets including deformation protocols beyond simple shear. Nevertheless, even under the limited shear dataset, the trained model stably and reliably reproduces the ground-truth tensorial stress response well (Appendix E).

\subsection*{Polydimethylsiloxane (PDMS)}
We next consider rheological data for a real complex fluid: a linear entangled homopolymer melt which is commonly used as a calibration standard for rheometers, PDMS. Small Amplitude Oscillatory Shear (SAOS) experiments were performed on an ARES-G2 rheometer (TA Instruments), and the discrete relaxation time spectrum was calculated from the SAOS data using the parsimonious fitting algorithm \cite{baumgaertel_determination_1989} implemented in the IRIS Rheo-Hub software \cite{winter_cyber_2006}. The linear viscoelastic spectrum measured over three decades of frequency (Fig.~\ref{fig:pdms}(a)) is compactly and accurately described by three discrete relaxation modes, with modal parameters (393.3 Pa, 0.4719 s), (8292 Pa, 0.06062 s), (51710 Pa, 0.007289 s). This gives a zero-shear rate viscosity $\eta_0 = 1065$ Pa and a viscosity-weighted relaxation time $\bar{\tau}_\eta \approx 0.113$ s. The mmRUDEs framework was subsequently trained on LAOS data obtained at an angular frequency of $\omega=2$~rad/s and four strain amplitudes $\gamma_0=1,2.5,4,5$, corresponding to Weissenberg numbers $\mathrm{Wi}=\bar{\tau}_\eta\,\omega\,\gamma_0 \approx 0.226,\ 0.565,\ 0.904,\ \text{and}\ 1.13$ (the normalization parameters used during training are $G_c = 9425 ~\text{Pa}$ and $\tau_c = 0.001 ~\text{s}$.). Only the shear stress was used in this training, i.e., the weight is $w_{N_1} = 0$ in the loss function. Within this trained window, the surrogate reproduces the experimental data almost exactly, as shown by the green shaded regime of Fig.~\ref{fig:pdms}(b). It also recovers the linear backbone at small strain amplitudes, and produces reasonable predictions up to strain amplitudes that are orders of magnitude greater than the values seen in the training regime. Accurate oscillatory test data at these large strain amplitudes are often inaccessible experimentally because of edge fracture or inertial instabilities that corrupt the data \cite{costanzo_review_2024}, hence the ability of the trained mmRUDE to produce robust and stable predictions in the large amplitude regime is useful. 

We next test the mmRUDE using unseen data obtained in the flow protocol corresponding to the start-up of steady shear (Fig. \ref{fig:pdms}(c)). Again, the trained model predicts the ground truth shear stress $\sigma_{12}(t;\dot{\gamma})$ behavior well, and is able to achieve good predictions even at high strain rates up to $\dot{\gamma} = 30$ s$^{-1}$ ($\mathrm{Wi} = 3.4$) that are outside of the training regime.

Once again, we can examine the contributions of each tensor basis $\bT{i}$ to the overall stress behavior predicted by the mmRUDE. We find that, similar to the Giesekus model, the $\bT{4}$ tensor contribution is the dominant correction learned by the mmRUDEs (Fig. \ref{fig:pdms}(e)), and our heat map representation shows that the non-zero 11, 22 and 12 terms increase in magnitude with shear rate (cf. Fig. \ref{fig:pdms}(e)). PDMS obeys the Cox-Merz rule \cite{cox_correlation_1958}, which enables an \textit{a priori} approximation of the steady shear viscosity from its linear viscoelastic spectrum. It also follows Laun's rule \cite{laun_prediction_1986,das_launs_2024}, which enables an \textit{a priori} estimation of the first normal stress difference under shear from the linear viscoelastic spectrum. The validity of these approximations is tested in Fig. \ref{fig:pdms}(d) and (f) (see hollow symbols). In both cases, we see excellent agreement at small $\mathrm{Wi}$ (as expected from the linear backbone of the RUDE) and, more remarkably, good agreement even up to values of the Weissenberg number that are more than an order of magnitude greater than those encountered in the training regime. This enhanced stability is significant, considering that previous implementations encountered numerical instabilities at moderate $\mathrm{Wi}$ only slightly beyond the training regime. The fact that the trained RUDE can produce physically reasonable predictions that remain close to scarce experimental data is a strong indication that this framework is learning a robust frame-indifferent constitutive representation that is physically meaningful for a real fluid. 
\newpage
\begin{figure*}[!t]
\centering
\includegraphics[width=\textwidth]{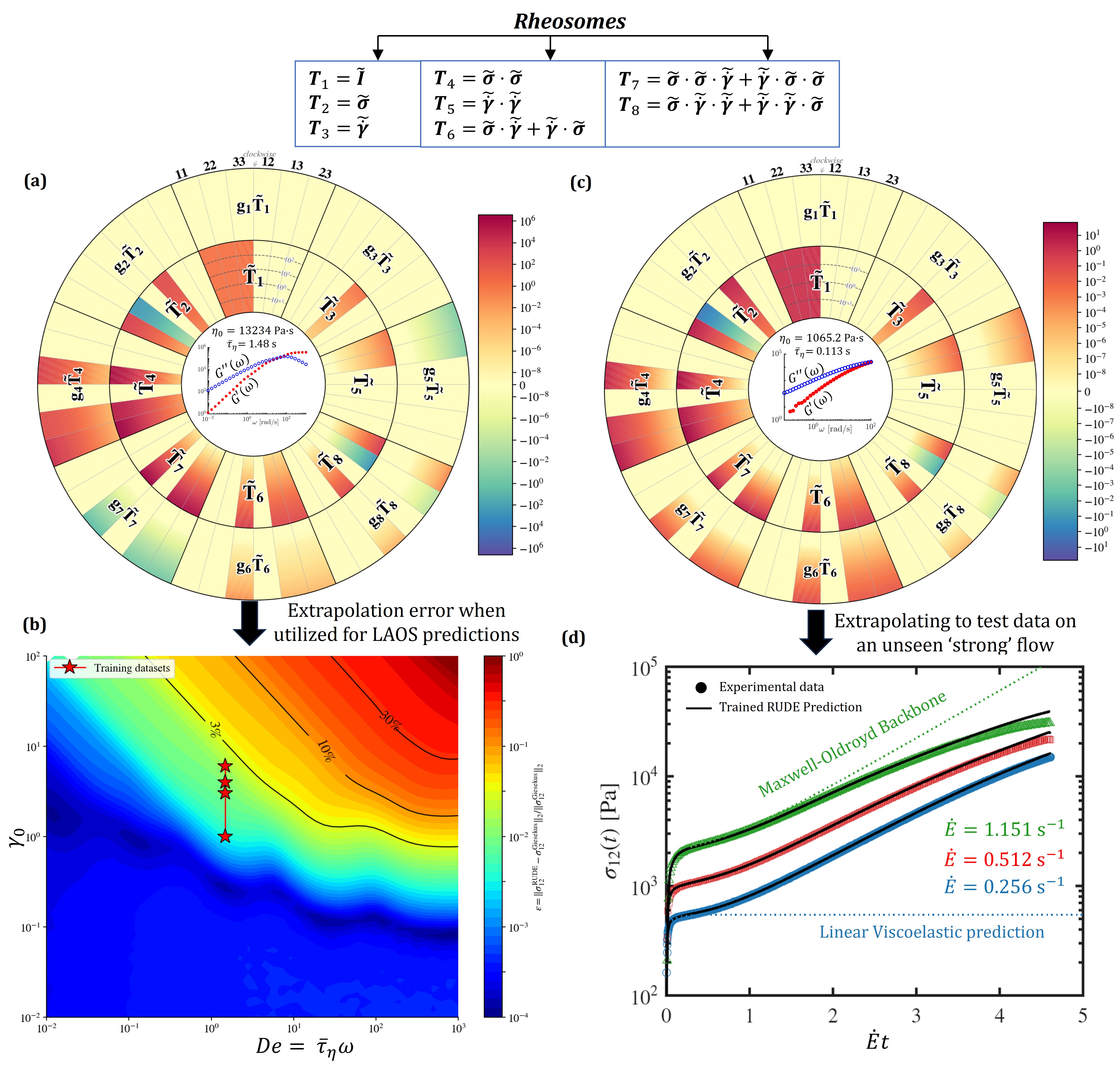}
\caption{\textbf{The \emph{Rheome}: an interpretable representation of a digital fluid twin through an mmRUDE.} (\textbf{a})~The \emph{rheome} of a learned surrogate model for a multi-mode Giesekus fluid. The radial map places the linear viscoelastic spectrum ($G'(\omega)$, $G''(\omega)$; center inset) at its core; the middle annulus shows the eight \emph{rheogenes} (i.e., the eight scaled basis tensors) $\tilde{\bT{1}}$ to $\tilde{\bT{8}}$, grouped into three \emph{rheosomes} by nonlinear order ($\{\tilde{\bT{1}}$, $\tilde{\bT{2}}$, $\tilde{\bT{3}}\}$, $\{\tilde{\bT{4}}$,$\tilde{\bT{5}}$,$\tilde{\bT{6}}\}$, $\{\tilde{\bT{7}}$,$\tilde{\bT{8}}\}$), and the outer annulus represents their expressed contribution $g_i\tilde{\bT{i}}$ towards measurable material properties. Each wedge is resolved into the six components of the symmetric tensor (clockwise) with deformation rate $\dot{\gamma} = \sqrt{\lambda_3/2}$ increasing logarithmically outwards by the indicated decadal spacing, and color encodes signed magnitude (symmetric-log colorbar). The learned expression concentrates on $g_4\tilde{\bT{4}}$, closely approximating the exact Giesekus closure $\bF=\alpha (\bsigma\cdot\bsigma)/G$. (\textbf{b})~Relative LAOS extrapolation error: red stars mark the four training
datasets, which is the sole nonlinear data provided. The error remains small ($\le 10\%$) over a broad region around the training conditions, growing only in the
strongest and most unsteady flows when $\mathrm{De\gg1}$, and $\mathrm{Wi}\gg 1$. (\textbf{c})~The learned \emph{rheome} of the entangled linear PDMS sample (same layout as (\textbf{a})). (\textbf{d})~Extrapolation for the trained mmRUDEs on PDMS data to an unseen strong flow history, exponential shear (Eq.~\ref{eq:exponential}), for
$\dot{E}=0.256,\,0.512,\,1.151~\mathrm{s}^{-1}$ (maximum $\mathrm{Wi}\approx13$). Symbols indicate experimental data, black lines are the trained mmRUDE predictions, the lower dotted line is the linear viscoelastic prediction for $\dot{E} = 0.256$ s$^{-1}$ and the upper dotted line is the prediction of the Maxwell Oldroyd B (MOB) backbone for $\dot{E} = 1.151$ s$^{-1}$.}
\vspace{12pt}
\label{fig:rheome}
\end{figure*}
\newpage

\section*{Interpretability of
  \texorpdfstring{\MakeLowercase{mm}RUDE\MakeLowercase{s}}{mmRUDEs}}
\label{sec:interpret}

Accurately reproducing the training data does not necessarily mean that the resulting tensorial model has identified the underlying precise constitutive relationship. This distinction is especially important for materials with unknown rheology. The key questions are (i) how much data is required to determine the underlying closure accurately, (ii) which flow protocols are sufficiently informative, and (iii) whether shear measurements alone are adequate, or whether extensional data plus other test protocols that excite distinct components of the stress tensor are also needed. To help answer such questions, it is essential to be able to visualize and quantify the learned constitutive response.

Motivated by prior work on genome map visualization, we propose the \emph{Rheome} as the rheological counterpart to a genome map \cite{krzywinski_circos_2009} for a complex fluid. We conceptualize each learned tensor basis function as carrying detailed constitutive information about a given complex fluid in the same way as genes carry quantitative genetic information about a given organism. These eight \emph{rheogenes} then represent the full tensorial output of the digital twin learned by the mmRUDE framework and collectively constitute the \emph{Rheome} as depicted in Fig.~\ref{fig:rheome}~(a,c). At its core is the linear viscoelastic spectrum, $G^*(\omega)$, the fundamental material response that is always experimentally accessible to a rheologist. Surrounding it are the eight \emph{rheogenes} $\bT{1}$ to $\bT{8}$, i.e., the admissible basis tensors, bundled into three \emph{rheosomes} by nonlinear order: the isotropic and linear terms $\{\bT{1},\bT{2},\bT{3}\}$, the quadratic terms $\{\bT{4},\bT{5},\bT{6}\}$, and the higher-order terms $\{\bT{7},\bT{8}\}$. The expression of each component of the tensorial \emph{rheogene} in a given flow history is encoded by its scalar function $g_i$, so the 6 radial subdivisions of each \emph{rheogene} segment in the \emph{rheome} quantify, in the form of heat maps, which stress components the training datasets activate. A small, sparse, and smoothly varying set of active \emph{rheogenes} defines a well-identified constitutive closure. 

The five-mode Giesekus fluid provides a clear test because its exact closure contains only the quadratic term, $\bsigma\cdot\bsigma$. The learned model reproduces this structure, activating almost exclusively the 11, 22 and 12 components of $g_4\bT{4}$ (Fig.~\ref{fig:rheome}(a)). Of the remaining functions, $g_1$ to $g_3\approx0$ and $g_5$ to $g_8$ are smaller, in most cases by orders of magnitude, although they are not exactly zero. The \emph{Rheome} confirms that one dominant, physically meaningful \emph{rheogene} has been successfully identified, with $g_7\bT{7}$ emerging as the second-largest contribution. Although $g_7(\lambda,\boldsymbol{\theta})$ is approximately four orders of magnitude smaller than $g_4(\lambda,\boldsymbol{\theta})$ within the validation window, the higher-order nature of the basis $\bT{7}=\boldsymbol{\sigma}\!\cdot\!\boldsymbol{\sigma}\!\cdot\!\dot{\boldsymbol{\gamma}}$ means that it grows more rapidly with shear rate. Consequently, $g_7\bT{7}$ becomes the most prominent residual contribution under strong flows (high Weissenberg numbers) and is therefore the \emph{rheogene} most likely to limit extrapolation beyond the validated regime. Here it stays negligible up to $\dot\gamma\simeq10^{5}~\mathrm{s}^{-1}~ (\mathrm{Wi} \simeq 1.48\times10^5)$, and the learned surrogate model continues to describe the Giesekus-like response accurately. Additional dimensional views of the individual components of the dominant \emph{rheogene} $g_4\bT{4}$ contribution are provided in Appendix~C, using both log--log and linear--linear axes. An analogous representation of the \emph{rheome} expressed in a uniaxial extensional flow is also shown, providing a complementary view of how the rheogenetic expression of the learned tensor-basis contributions reorganize under a different deformation protocol (see Appendix~C, Fig.~\ref{fig:giesekus_si}). The extrapolation-error map of Fig.~\ref{fig:rheome}(b) quantifies the learned performance of the digital twin. Even though the surrogate model is only trained at a single Deborah number (which provides a measure of flow unsteadiness), corresponding to $De=\bar\tau_\eta\omega = 1.48$ and four strain amplitudes $\gamma_0$ (vertical axis), the relative error 
\[
\epsilon
=
\frac{
\left\|
\sigma_{12}^{\mathrm{RUDE}}
-
\sigma_{12}^{\mathrm{Giesekus}}
\right\|_2
}{
\left\|
\sigma_{12}^{\mathrm{Giesekus}}
\right\|_2
}
\] 
stays small over a broad region around the training conditions and rises only in the strongest flows. For large-amplitude oscillatory flows, an increase in Weissenberg number  ($\mathrm{Wi} =\bar{\tau}_\eta\,\gamma_0\,\omega$) is a diagonal movement towards the top right of this map, which is where the largest errors of $\geq 30\%$ are incurred. 

The visual interpretability provided by our \textit{rheome} representation also applies for digital twins of a real complex fluid. Fig.~\ref{fig:rheome}(c) shows the \emph{rheome} learned for a linear entangled PDMS homopolymer melt, for which the exact constitutive closure is not known \textit{a priori}. The model is trained on a limited set of shear measurements that depart only weakly from the linear regime. The first normal stress difference signal at low Weissenberg numbers is degraded by experimental noise and cannot be reliably measured. It is thus essential that reliable training can be performed using sparse data availability. It is evident that the dominant terms in the \emph{rheome} can still be clearly identified, and the model again extrapolates stably across the previously unseen transient flow startup tests presented in Fig.~\ref{fig:pdms}, indicating that the available data were still sufficient to identify a robust and accurate approximation of the underlying constitutive equation.

To further illustrate the performance of the mmRUDE framework on a kinematically distinct (and unseen) flow that is strongly nonlinear, we show in Fig.~\ref{fig:rheome}(d), data for PDMS subject to a time-dependent exponential shearing flow \cite{doshi_exponential_1987}. This is a kinematically strong flow that can be performed in a commercial rheometer by applying a strain rate that is exponentially increasing with time: 
\begin{equation}
    \dot{\gamma} (t) = 2\dot{E} \cosh(\dot{E} t).
    \label{eq:exponential}
\end{equation} 
Here $\dot{E}$ is a parameter that controls the rate of increase of the strain rate and the product $\dot{E} t$ is a measure of the total Hencky strain. In this protocol material points separate exponentially in time, and the fluid is rapidly stretched at high Weissenberg numbers. For $\dot{E} = 1.151$ s$^{-1}$, at the end of the experiment ($t = 4$ s), the Hencky strain is $\dot{E} t = 4.5$, and the maximum value of $\mathrm{Wi} = 13$, which is an order of magnitude greater than any Weissenberg number observed during training. Nevertheless, the trained mmRUDE can produce very reasonable predictions even when extrapolated far beyond the training regime, as shown in Fig.~\ref{fig:rheome}(d). The linear viscoelastic response due to the initial imposed step strain rate (for $\dot{E} = 0.256$ s$^{-1}$) and the expected MOB response (i.e. the model prediction if no constitutive closure is learned, $\bF = \pmb{0}$) at the highest imposed deformation rate are also plotted as dotted lines for comparison, and these \textit{a priori} predictions bound the actual measured data and the predicted material response.

To summarize, our \emph{Rheome} representation helps us organize and visualize the strength of every component of each basis tensor (i.e. the complex fluid's \emph{rheogenes}) expressed in the mmRUDE under steady shear flow. This allows for the simple identification of the dominant tensor basis functions. Analogous heatmaps can be constructed for other flows, such as uniaxial extension, where the \emph{rheogenes} will be expressed through different non-zero tensor components; since the coefficients $g_i$ in Eq. \ref{eq:tbnn} depend only on the flow invariants, a reliably trained model recovers the same dominant closure regardless of the flow type.

\section*{Discussion}

In this work, we have focused on learning compact and robust constitutive equations for polymeric liquids, which are perhaps the most well-understood subclass of complex fluids \cite{larson_structure_1999}. Our proposed thermodynamic constraint, based on a Hookean dumbbell model, is also the least restrictive among polymer models because it can be applied to both the dilute solution of dumbbells used to model polymer solutions and also to transient junctions within a polymer network in a more concentrated polymer melt \cite{larson_constitutive_1988}. More complicated closures such as the FENE-P \cite{bird_polymer_1980} spring law can also be incorporated through our formulation of the thermodynamic constraint. However, because the constraint becomes non-affine in the stress term, a more general version of HardNet, known as HardNet++ \cite{goertzen_hardnet_2026}, should be applied in this instance. For other subclasses of complex fluids, there may be other more suitable formulations of physically-based constraints depending on the specific microstructure being represented, and the HardNet projection provides a simple way to incorporate these constraints exactly into the neural architecture. 

Although deployed here in the context of UDEs, the HardNet projection can also serve as a post-processing step for other methods to discover constitutive equations. For example, sparse regression methods may produce a constitutive equation that violates thermodynamic constraints \cite{shanbhag_sparse_2024}. Projecting the learned equation onto an admissible set can yield the closest thermodynamically consistent model (with respect to the metric defined by the HardNet projection). This projection therefore imposes physical admissibility while preserving the compactness and interpretability of the original equation learned from sparse regression. The usefulness of the HardNet projection layer therefore extends beyond mmRUDE training to a more general refinement procedure for learned constitutive models of complex fluids. 

It is well known that the representation of an unknown material's relaxation spectrum by a discrete Prony series is not unique \cite{bird_dynamics_1987}. Nevertheless, Shanbhag \cite{shanbhag_does_2025} has shown that this non-uniqueness is not generally a limitation for representing an unknown material's stress response. Any compact set of discrete relaxation modes that can reproduce the observable linear viscoelastic data of a material can be used for the backbone of the mmRUDE. The trained mmRUDE can then still produce robust descriptions of an unknown material, as we have illustrated with the PDMS example in Fig. \ref{fig:pdms}.

More fundamentally, the co-linearity of several of the eight tensor bases in homogeneous shearing deformations (as well as in analogous homogeneous extensional flow) is a limitation to learning the most general constitutive equation for a complex fluid. Models trained on shear data alone will remain non-unique with respect to the individual scalar functions $g_i(\lambda_i;\theta)$ even with our thermodynamic constraint. Fully unique identification of all eight scalar functions requires mixed flow protocols that excite all six independent components of the tensorial correction $\bF$ either simultaneously, or sequentially. However, such flows typically combine both shear and extensional components, and spatial homogeneity will be harder to satisfy and measurement complexity increases, although some hypothetical rheometer designs have been proposed \cite{dayal_design_2012,giusteri_theoretical_2018}. Access to larger portions of the kinematic parameter space (in terms of the deformation gradient tensor) is possible when interest is directed towards the data-driven constitutive modeling of solids \cite{fuhg_review_2025}: beyond simple shear and homogeneous extension, solid materials can also be subject to more complex deformations such as multi-axial loading, combined tension and torsion, or more recently, even inhomogeneous strain fields which can be measured by digital image correlation \cite{pierron_towards_2021}. Similar diversity in the types of flows used to probe complex fluids is difficult to achieve in a rheometer, however, bespoke devices to generate more varied flows are being actively developed by the rheological community \cite{galindo-rosales_microdevices_2013}. In principle, any type of rheometric data, provided the kinematic history is carefully imposed and quantified, can be used to generate training data for our mmRUDE formulation. Future work should focus on incorporating novel training protocols to expand the robustness and extrapolation capabilities of the learned digital surrogate. Because of the large number of parameters in the neural network embodied by Fig. \ref{fig:architecture}, any additional data used to train the model will help refine the model weights and biases to improve the neural network prediction capability. The robustness of our multi-mode Maxwell--Oldroyd-B backbone, coupled with the design of the two neural networks, and the stability conferred by the thermodynamic HardNet projection layer ensure that this data will be stably integrated into refinement of the rheogenetic information embodied in the \emph{rheomes} identified in Figs. \ref{fig:rheome}(a) and (c). Ideally, the design of new rheometric tests should develop concurrently with the data demands of mmRUDE training protocols, and such a synergistic approach will be fruitful for the development of more predictive constitutive equations.

\section*{Conclusions}

This work advances data-driven constitutive modeling toward learned material models of complex fluids that are not only accurate, but also stable, physically admissible, and interpretable. We have developed a multi-mode Rheological Universal Differential Equation framework (mmRUDEs), for polymeric materials with broad relaxation spectra by embedding neural-network flexibility within a frame-indifferent constitutive architecture. The framework combines (i) log-modulus compression of the input invariants, (ii) a split-neural network structure that separates low-order and higher-order tensor-basis function contributions, and (iii) a differentiable projection layer that enforces thermodynamic admissibility of the learned correction. Together, these features convert the original flexible, but fragile, data-driven closure into a robust multi-mode constitutive surrogate that preserves the measured linear viscoelastic response while learning nonlinear rheological behavior of an unknown material from limited data.

The trained mmRUDEs remain accurate and numerically stable beyond the deformation conditions used for training and provide robust and accurate predictions under previously unseen flow protocols. In synthetic benchmarks, where the underlying true frame-invariant constitutive equation is known, the framework recovers the expected dominant nonlinear tensor-basis contribution. When trained against experimental data for an entangled polymer melt, it produces stable and physically meaningful predictions in nonlinear flow regimes that extend well beyond the original training window. These results demonstrate that incorporating rheological and thermodynamic admissibility directly into scientific machine learning can substantially improve the robustness of data-driven constitutive models.

We also introduce the \emph{rheome} as a compact visual representation of the learned constitutive closure. By organizing the active tensor-basis contributions (which one may consider as the underlying \emph{rheogenes} governing the material response characteristics), the \emph{rheome} makes the nonlinear response of the trained digital fluid twin directly inspectable rather than remaining hidden within a black-box neural network. This added interpretability is useful for assessing what has been learned, identifying dominant constitutive mechanisms, and guiding the design of more informative rheological experiments.

More broadly, mmRUDEs provide a promising route from sparse rheological measurements to transferable digital representations of complex fluids. By unifying stability, thermodynamic consistency, predictive accuracy, and interpretability, this framework lays a foundation for automated learning of constitutive equations that can support predictive flow simulations and the rational design and control of soft-material processing operations.
$\bT{4}=\bsigma\cdot\bsigma$, with a coefficient close to the analytic value—identifying the correct constitutive equation. We then considered developing a digital fluid twin for an entangled polymer melt and showed that the resulting closure is able to produce predictions in unseen flow protocols and to higher Weissenberg numbers than experimentally accessible.

\section*{Acknowledgements}
We sincerely thank Eugene Pashkovski (Lubrizol) for sharing his experimental PDMS data. Financial support from 3M (Damian C. Vadillo and Alessandro Perego) and the Lubrizol Corporation (Eugene Pashkovski and Reid Patterson), as well as helpful discussions regarding the industrial requirements and formulation of digital fluid twins are gratefully acknowledged.

\section*{Data, Materials, and Software Availability}
All code and data used during preparation of this manuscript are available from the corresponding author upon reasonable request, and will eventually be made available on GitHub upon publication of the manuscript. 

\section*{Competing interests} The authors declare no competing interests.

\section*{Author contributions}
MD, NK, GHM designed the research; MD, NK performed the research; NA contributed     new analytic tools; MD, NK analyzed data; MD, NK, NA and GHM wrote and edited the paper.

\printbibliography

\onecolumn
\newpage

\appendix
\section*{SI Appendix A: Derivation of the Thermodynamic Projection Layer}
\addcontentsline{toc}{section}{SI Appendix A}

\setcounter{equation}{0}
\renewcommand{\theequation}{A\arabic{equation}}

\subsection{Second Law of Thermodynamics}

It is well-known in continuum mechanics that for a spatially and temporally isothermal, incompressible system, the local form of the second law of thermodynamics (Clausius-Duhem inequality), which is a free-energy imbalance \cite{anand_continuum_2020}, is stated in the following way:
\begin{equation}
  \rho \dot\Psi - \bm{T}:\nabla\bm{v} \;\leq\; 0.
  \label{eq:si_CD_full}
\end{equation}

in which $\rho$ is the density of the material, $\dot\Psi$ is the rate of change of the Helmholtz free energy per unit mass, $\bm{T}$ is the total stress tensor, and $\nabla \bm{v}$ is the velocity gradient tensor. Note that in some texts, the expression may be written in a different way because $\bm{T}$ is symmetric, so $\bm{T} : \nabla \bm{v} = \bm{T} : \bm{D}$, in which $\bm{D} = \tfrac{1}{2}(\nabla \bm{v} + (\nabla \bm{v})^T)$ is the symmetric part of the velocity gradient tensor.

When modeling complex fluids, the total stress tensor is often decomposed as $\bm{T} = -p \bm{I} + \bsigma$, in which $p$ is the isotropic pressure term and $\bsigma$ is the deviatoric extra stress tensor that is related to the rate-of-strain tensor $\dot{\pmb{\gamma}} = 2\bm{D}$ through a constitutive relation. Because $\bm{D}$ is traceless (due to incompressibility), the inequality can also be written in terms of $\bsigma$:
\begin{equation}
  \rho \dot\Psi - \bsigma:\nabla\bm{v} \;\leq\; 0.
  \label{eq:si_CD}
\end{equation}

For the specific form of the Helmholtz free energy, some assumptions must be made about the microstructure of the complex fluid. We consider a dilute polymer solution modeled as a suspension of elastic dumbbells. The configuration of each dumbbell is described by its conformation tensor $\pmb{c}$, which is a symmetric positive-definite rank-2 tensor. The Helmholtz free energy per unit volume is hence \cite{nieto_simavilla_hammering_2025}:
\begin{equation}
    \rho\,\Psi = \frac{G}{2}\Bigl[
      \Phi\bigl(\mathrm{tr}\,\pmb{c}\bigr)
      - \ln\det\pmb{c} - 3
    \Bigr],
  \label{eq:si_free_energy}
\end{equation}
where $G = n k_B T$ is the elastic modulus, which depends on the number density of dumbbells $n$ and the temperature $T$, and $\Phi(\mathrm{tr}\, \pmb{c})$
is a spring potential. The dimensionless spring function $f$ is defined as the derivative of the spring potential:
\begin{equation}
  f\bigl(\mathrm{tr}\,\pmb{c}\bigr)
  \;\equiv\;
  \frac{\dd\Phi}{\dd(\mathrm{tr}\,\pmb{c})}
  \label{eq:si_spring_function}
\end{equation}

The stress tensor $\bsigma$ is related to the conformation tensor $\pmb{c}$ through the following relation:
\begin{equation}
  \bsigma = G\bigl(f\, \pmb{c} - \bI\bigr),
  \label{eq:si_stress_conf}
\end{equation}

To obtain the rate of change of Helmholtz free energy, we differentiate Eq. \ref{eq:si_free_energy} with respect to time:
\begin{equation}
    \rho\,\dot\Psi = \frac{G}{2}\left[
      \frac{\dd}{\dd t}\Phi\bigl(\mathrm{tr}\,\pmb{c}\bigr)
      - \frac{\dd}{\dd t}\ln\det\pmb{c}
    \right]
    \label{eq:si_free}
\end{equation}

The first term can be found by the chain rule:
\begin{align}
  \frac{\dd}{\dd t}\Phi
  = \frac{\dd\Phi}{\dd(\mathrm{tr}\,\pmb{c})}\cdot
     \frac{\dd}{\dd t}\mathrm{tr}(\pmb{c})
  = f\;\mathrm{tr}\!\left(\frac{\dd \pmb{c}}{\dd t}\right)
  = f\;\mathrm{tr}\bigl(\dot{\pmb{c}}\bigr)
  = f\,\bI : \dot{\pmb{c}}
  \label{eq:si_first_term}
\end{align}

The second term can be found by applying Jacobi's formula for the derivative of a determinant in the second step below:
\begin{align}
    \frac{\dd}{\dd t}\ln\det \pmb{c}
    = \frac{1}{\det\pmb{c}}\;\frac{\dd}{\dd t}\bigl[\det\pmb{c}\bigr] 
    = \frac{1}{\det\pmb{c}}\;\det\pmb{c} \; \mathrm{tr} \left(\pmb{c}^{-1}\dot{\pmb{c}} \right) 
    = \mathrm{tr} \left(\pmb{c}^{-1}\dot{\pmb{c}} \right) 
    = \pmb{c}^{-1} : \dot{\pmb{c}}
\end{align}

This expression \ref{eq:si_free} is hence simplified:
\begin{align}
    \rho\,\dot\Psi &= \frac{G}{2}\left[
      f\,\bI : \dot{\pmb{c}} - \pmb{c}^{-1} : \dot{\pmb{c}}
    \right] \\
    &= \frac{G}{2} \bigl(f\,\bI - \pmb{c}^{-1}\bigr):\dot{\pmb{c}}
\end{align}

For the \textbf{specific case of a Hookean spring}, $f = 1$. This is the least restrictive assumption and applies to the Oldroyd-B and Giesekus models used in this work. The stress-conformation relation then simplifies to:
\begin{equation}
    \bsigma = G( \pmb{c} - \bI)
\end{equation}
\begin{equation}
    \dot{\bsigma} = G \dot{\pmb{c}}
\end{equation}

From this point onward, the derivation specializes to the limit \(f=1\), so \(f\) is constant and \(\dot{\boldsymbol{\sigma}}=G\dot{\mathbf{c}}\); for non-Hookean spring laws, differentiating \(\boldsymbol{\sigma}=G(f\mathbf{c}-\mathbf{I})\) introduces the additional term \(G\dot{f}\,\mathbf{c}\).

The rate of change of stress is also linked to the learned RUDE expression:
\begin{equation}
  \dot{\bsigma}
  = \nabla\bm{v}\!\cdot\!\bsigma
  + \bsigma\!\cdot\!(\nabla\bm{v})^{\!T}
  + G\,\bgdot
  - \frac{\bsigma + \bF}{\tau}.
  \label{eq:si_sigma_dot_explicit}
\end{equation}

The above can now all be substituted into the free energy imbalance:
\begin{align}
    \rho \dot\Psi - \bsigma:\nabla\bm{v} \;&\leq\; 0 \\
    \frac{G}{2} \bigl(\bI - \pmb{c}^{-1}\bigr):\dot{\pmb{c}} - \bsigma:\nabla\bm{v} \;&\leq \; 0 \\
    \frac{1}{2} \bigl(\bI - \pmb{c}^{-1}\bigr):\dot{\bsigma} -  \bsigma:\nabla\bm{v} \;&\leq \; 0 \\
    \frac{1}{2} \bigl(\bI - \pmb{c}^{-1}\bigr):\left[
       \nabla\bm{v}\!\cdot\!\bsigma
       + \bsigma\!\cdot\!(\nabla\bm{v})^{\!T}
       + G\,\bgdot
       - \frac{\bsigma + \bF}{\tau}
     \right] -  \bsigma:\nabla\bm{v} \;&\leq \; 0 \\
     \underbrace{
       \frac{1}{2}\bigl(\bI - \pmb{c}^{-1}\bigr)
       :\bigl(
         \nabla\bm{v}\!\cdot\!\bsigma
         + \bsigma\!\cdot\!(\nabla\bm{v})^{\!T}
         + G\,\bgdot
       \bigr)
     }_{\text{Term A }}
     \;-\;
     \underbrace{
       \frac{1}{2\tau}\bigl(\bI - \pmb{c}^{-1}\bigr)
       :(\bsigma + \bF)
     }_{\text{Term B (relaxation $+$ correction)}} -  \bsigma:\nabla\bm{v} \;&\leq \; 0
\end{align}

Term A can be broken down into two parts. For the $\bI$ part, since $\bsigma$ is symmetric and $\mathrm{tr}(\bgdot) = 0$ (incompressibility):
\begin{align}
  \bI:\bigl(
    \nabla\bm{v}\!\cdot\!\bsigma
    + \bsigma\!\cdot\!(\nabla\bm{v})^{\!T}
    + G\,\bgdot
  \bigr)
  &= \mathrm{tr}(\nabla\bm{v}\!\cdot\!\bsigma)
   + \mathrm{tr}(\bsigma\!\cdot\!(\nabla\bm{v})^{\!T})
   + G\,\mathrm{tr}(\bgdot)
  \notag\\[4pt]
  &= 2\,\bsigma:\nabla\bm{v}
\end{align}

For the $\pmb{c}^{-1}$ part, substituting $\bsigma = G(\pmb{c} - \bI)$ to expand each sub-term:
\begin{align}
  \pmb{c}^{-1}:(\nabla\bm{v}\!\cdot\!\bsigma)
  &= G\bigl[
       \underbrace{
         \pmb{c}^{-1}:(\nabla\bm{v}\!\cdot\!\pmb{c})
       }_{=\,\mathrm{tr}(\nabla\bm{v})=0}
       -\;
       \pmb{c}^{-1}:\nabla\bm{v}
     \bigr]
  = -\,G\;\pmb{c}^{-1}:\nabla\bm{v},
  \\[6pt]
  \pmb{c}^{-1}:(\bsigma\!\cdot\!\nabla\bm{v}^{\!T})
  &= G\bigl[
       \underbrace{
         \pmb{c}^{-1}:(\pmb{c}\!\cdot\!(\nabla\bm{v})^{\!T})
       }_{=\,\mathrm{tr}(\nabla\bm{v}^T)=0}
       -\;
       \pmb{c}^{-1}:(\nabla\bm{v})^{\!T}
     \bigr]
  = -\,G\;\pmb{c}^{-1}:(\nabla\bm{v})^{\!T},
  \\[6pt]
  \pmb{c}^{-1}:G\,\bgdot
  &= G\;\pmb{c}^{-1}:(\nabla\bm{v} + (\nabla\bm{v})^{\!T}).
\end{align}

Hence, the $\pmb{c}^{-1}$ part sums up to zero. Adding up these four contributions, Term A has a simple form:

\begin{equation}
  \text{Term A}
    = \tfrac{1}{2}\bigl[2\,\bsigma:\nabla\bm{v} - 0\bigr]
    = \bsigma:\nabla\bm{v}.
  \;
  \label{eq:si_termA}
\end{equation}

Term B cannot be simplified further. Hence, we obtain the following inequality:
\begin{align}
  \cancel{\bsigma:\nabla\bm{v}}
  - \frac{1}{2\tau}\bigl(\bI - \pmb{c}^{-1}\bigr):(\bsigma + \bF)
  - \cancel{\bsigma:\nabla\bm{v}}
  &\;\leq\; 0
\end{align}

The final constraint has a simple form:
\begin{equation}
  \boxed{\;
  (f\,\bI - \pmb{c}^{-1}):(\bsigma + \bF) \;\geq\; 0}
  \label{eq:si_dissipation_constraint}
\end{equation}

For classical models in which the nonlinear term depends only on stress (e.g. for the Giesekus model, $\bF \propto \bsigma\!\cdot\!\bsigma$, and for the linear and exponential PTT model, $\bF \propto \bsigma$), this condition is automatically satisfied. Models that include a Gordon-Schowalter slip parameter $\xi$, such as the Johnson-Segalman model ($\bF \propto \bgdot\!\cdot\!\bsigma + \bsigma\!\cdot\!\bgdot$), may violate it in extreme flow states, which indicates a thermodynamic inconsistency in the model for some parameters \cite{larson_constitutive_1988}. The tensor basis is expressive enough to represent these corrections (the slip maps exactly onto $\bT{6}$), and in standard rheometric protocols the backbone dissipation dominates so the constraint is inactive. The projection becomes active only in the regime where the unconstrained surrogate would diverge. The net effect is a stable surrogate that matches the target model in the training regime and remains deployable where the unconstrained model would fail. The thermodynamic constraint narrows the space of admissible solutions. Without it, many combinations reproduce the same shear data in the training regime equally well, and the optimizer has no reason to prefer the compact, physically correct one.

\subsection*{Minimum-Norm Projection}

Without implementing this hard thermodynamic constraint, the RUDE produces a tensor $\bF_\mathrm{raw}$, which may or may not violate the constraint. The concept of hard-constrained neural nets (HardNets) proposed by Min and Azizan \cite{min_hardnet_2025} is to add a projection layer at the end of the neural network that takes the raw output and either (1) does not modify it if the constraint is satisfied, or (2) projects the output onto the boundary of the feasible half-space if the constraint is violated. The projection formula is:

\begin{equation}
  \boxed{\bF_\mathrm{proj}
  = \bF_\mathrm{raw}
  + \frac{(f\,\bI - \pmb{c}^{-1})}{\|(f\,\bI - \pmb{c}^{-1})\|^2}\;
    \texttt{ReLU}\!\Bigl(-(f\,\bI - \pmb{c}^{-1}):(\bsigma + \bF_\mathrm{raw})\Bigr),
  \quad
  \|(f\,\bI - \pmb{c}^{-1})\|^2 = (f\,\bI - \pmb{c}^{-1}):(f\,\bI - \pmb{c}^{-1})}
  \label{eq:si_projection}
\end{equation}

We illustrate how the projection layer works for two cases:

\textbf{Case~I: Constraint satisfied}
($(f\,\bI - \pmb{c}^{-1}):(\bsigma + \bF_\mathrm{raw}) \geq 0$).

The argument of the $\texttt{ReLU}$ function is $\leq 0$, hence $\texttt{ReLU}\!\bigl(-(f\,\bI - \pmb{c}^{-1}):(\bsigma + \bF_\mathrm{raw})\bigr) = 0$. This leads to the simple statement that $\bF_\mathrm{proj} = \bF_\mathrm{raw}$, essentially indicating that no correction is applied.

\textbf{Case~II: Constraint violated}
($(f\,\bI - \pmb{c}^{-1}):(\bsigma + \bF_\mathrm{raw}) < 0$).

The argument of the $\texttt{ReLU}$ function is $\geq 0$, hence $\texttt{ReLU}\!\bigl(-(f\,\bI - \pmb{c}^{-1}):(\bsigma + \bF_\mathrm{raw})\bigr) = -(f\,\bI - \pmb{c}^{-1}):(\bsigma + \bF_\mathrm{raw})$. We can verify that the projected $\bF$ lies on the boundary. The projected $\bF$ is:
\begin{equation}
  \bF_\mathrm{proj}
  = \bF_\mathrm{raw}
  + \frac{-(f\,\bI - \pmb{c}^{-1}):(\bsigma + \bF_\mathrm{raw})}{\|(f\,\bI - \pmb{c}^{-1})\|^2}\;(f\,\bI - \pmb{c}^{-1})
\end{equation}

We can check whether this satisfies the constraint:
\begin{align}
  (f\,\bI - \pmb{c}^{-1}):(\bsigma + \bF_\mathrm{proj})
  &= (f\,\bI - \pmb{c}^{-1}):\bsigma + (f\,\bI - \pmb{c}^{-1}):\bF_\mathrm{raw}
     + \frac{-(f\,\bI - \pmb{c}^{-1}):(\bsigma + \bF_\mathrm{raw})}{\|(f\,\bI - \pmb{c}^{-1})\|^2}\;((f\,\bI - \pmb{c}^{-1}):(f\,\bI - \pmb{c}^{-1}))
  \notag\\[4pt]
  &= 
       (f\,\bI - \pmb{c}^{-1}):(\bsigma + \bF_\mathrm{raw}) -(f\,\bI - \pmb{c}^{-1}):(\bsigma + \bF_\mathrm{raw})
  \notag\\[4pt]
  &= 0
\end{align}
This indicates that the constraint is satisfied minimally.

\newpage
\appendix
\section*{SI Appendix B: Ablation: the projection shapes learned weights}
\addcontentsline{toc}{section}{SI Appendix B}
To distinguish whether stability is due to the learned weights or from the thermodynamic projection operator being applied during the inference stage, we cross-evaluate both weight sets with the thermodynamic projection layer enabled and disabled at both training and extrapolation. Weights trained with the thermodynamic projection layer produce stable predictions even with the projection turned off, while weights trained without the thermodynamic projection layer still diverge even when the projection is enabled at inference. The dominant benefit of the thermodynamic layer is hence not a post-hoc runtime fix: the thermodynamic constraint steers the training of the neural network toward a different, more stable region of weight space.

\begin{figure*}[!htbp]
\centering
\includegraphics[width=\textwidth]{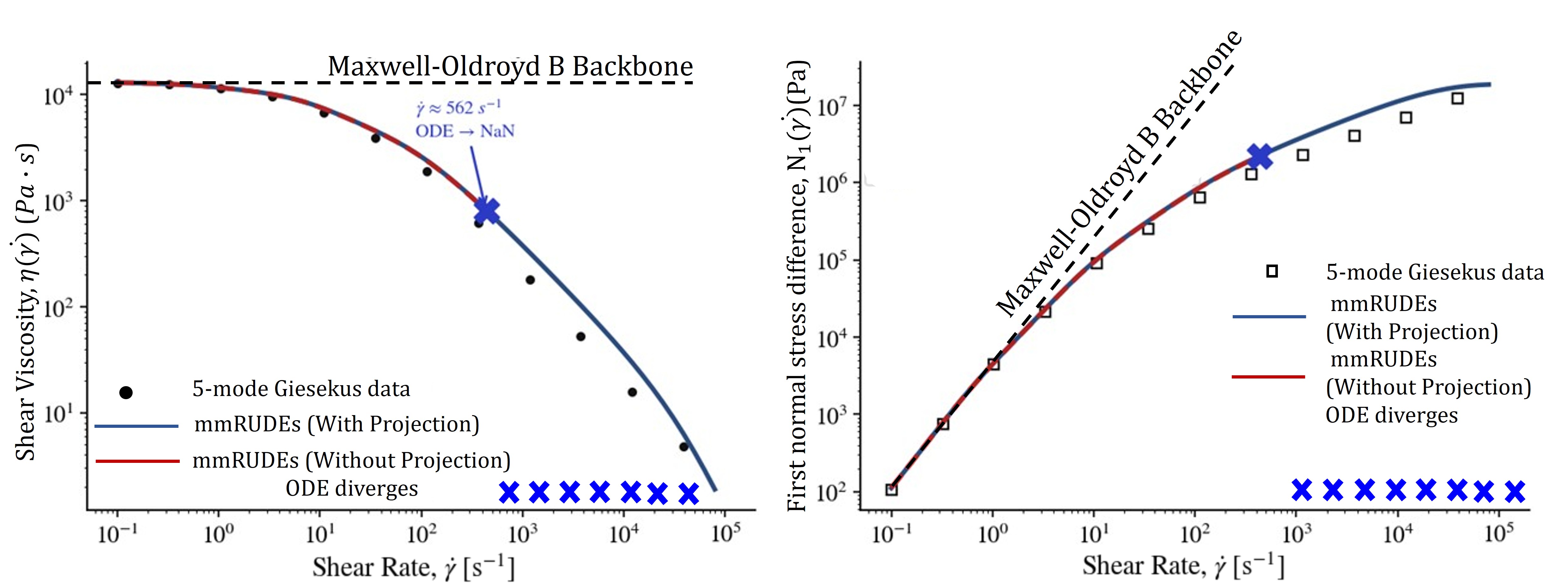}

\caption{\textbf{Ablation study: stability under unseen steady
shear.}
Steady-state shear viscosity $\eta(\dot\gamma)$ (\textit{left}) and first
normal stress difference $N_1(\dot\gamma)$ (\textit{right}) versus shear
rate $\dot{\gamma}$.
Symbols: five-mode Giesekus ground truth;
blue solid line: mmRUDEs with thermodynamic projection;
red dashed line: mmRUDEs without projection.
The unprojected model diverges at
$\dot{\gamma}\approx 562~\mathrm{s}^{-1}$ (crosses mark
blow-up), while the projected model remains stable up to
$\dot{\gamma} = 10{,}000~\mathrm{s}^{-1}$ and beyond.
Steady shear was \emph{never} included in the training data.}
\label{fig:ablation}
\end{figure*}

We evaluate both the projection-trained and no-projection models on steady shear across six decades of shear rate ($\dot{\gamma} = 0.1$--$100{,}000~\mathrm{s}^{-1}$), an entirely unseen flow protocol from the training data. The model trained without projection diverges at $\dot{\gamma} \approx 562~\mathrm{s}^{-1}$, with the ODE solver returning NaN; for the slowest mode ($\tau_1 = 100$~s), this shear rate corresponds to $\Wi_1 \approx 5.6\times 10^4$, well outside the oscillatory-training envelope. 
The projection-trained model remains stable over the full four-decade range, predicting both the steady-shear viscosity $\eta(\dot{\gamma}) \equiv \sigma_{12}/\dot{\gamma}$ and the first normal-stress difference in agreement with the ground truth (Fig.~\ref{fig:ablation}).

\newpage
\appendix
\section*{SI Appendix C: Further Results on the 5-mode Giesekus Model}

\begin{figure*}[!htbp]
\centering
\includegraphics[width=\textwidth]{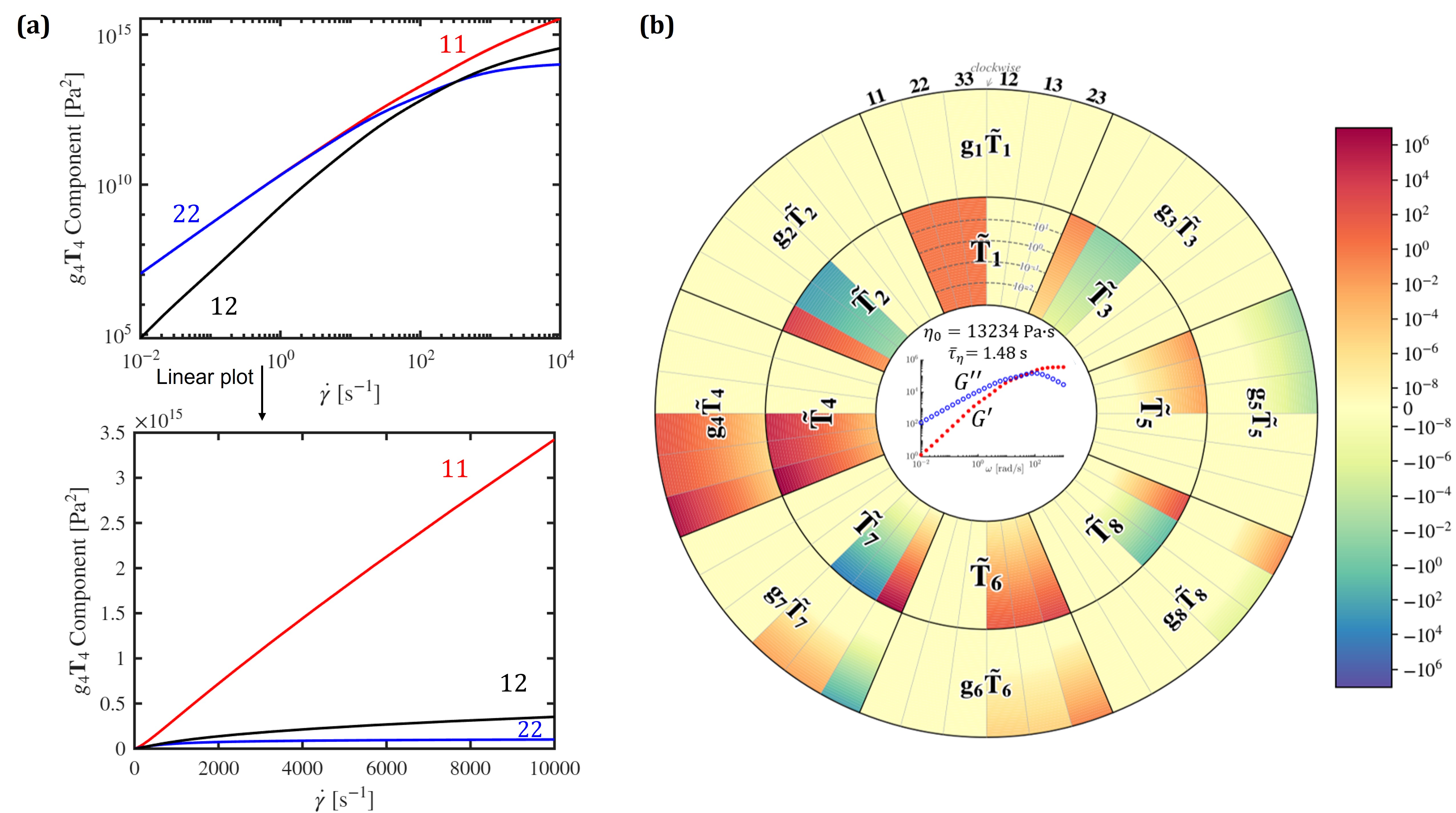}
\caption{
\textbf{Further results for the five-mode Giesekus model. }
(a) Redimensionalized $11$, $22$, and $12$ components of the dominant \emph{rheogene} $g_4\mathbf{T}_4$ under steady shear (presented in Fig. \ref{fig:training giesekus}(a) of main test) as functions of shear rate, shown on logarithmic and linear axes. 
(b) \emph{Rheome} of the trained mmRUDE under steady uniaxial extension. 
Model parameters are the same as those used in Fig.~\ref{fig:rheome} (a) of the main text.
}
\label{fig:giesekus_si}
\end{figure*}

Further information on the trained mmRUDE for the five-mode Giesekus model is presented in Fig.~\ref{fig:giesekus_si}. Fig.~\ref{fig:giesekus_si}(a) provides a dimensional view of the three non-zero components of the dominant \emph{rheogene}, $g_4\mathbf{T}_4$, identified in the steady-shear \emph{Rheome} of Fig.~\ref{fig:rheome} (a). The redimensionalized $11$, $22$, and $12$ components are shown as functions of shear rate on both logarithmic and linear axes. This representation highlights the quantitative numerical differences between the $11$ and $22$ components, which then generate the contribution of this \emph{rheogene} at high shear rates to the first normal stress difference, $g_4 (T_{4,11} - T_{4,22})$, while the $12$ component, $g_4 T_{4,12}$, contributes directly to the shear stress.

Fig.~\ref{fig:giesekus_si}(b) shows the \emph{Rheome} obtained from predictions of the same trained mmRUDE under steady uniaxial extension. The overall structure remains qualitatively similar to that obtained in steady shear, reflecting the same learned scalar functions $g_1$--$g_8$, while their tensorial contributions change with the imposed flow kinematics. This provides an additional view of how the constitutive structure learned from the training data is expressed under a different deformation protocol.

\newpage
\appendix
\section*{SI Appendix D: Performance Evaluation: Multi-mode Giesekus Model with different $\alpha$}

The linear viscoelastic spectra is described by two modes: (1000 Pa, 1 s), (5000 Pa, 0.001 s). The first mode has Giesekus mobility parameter $\alpha_1 = 0.1$ and the second mode has $\alpha_2 = 0.05$. The viscosity-weighted relaxation time is $\bar{\tau}_\eta = 0.995$ s. The trained mmRUDE is able to reproduce the ground truth data well, despite the use of a common nonlinear term learned by the mmRUDE. This implies that the mmRUDE has the flexibility to learn a suitable correction term that may not be exactly the ground truth, yet can produce reasonable predictions. 

\begin{figure*}[!htbp]
\centering
\includegraphics[width=0.8\textwidth]{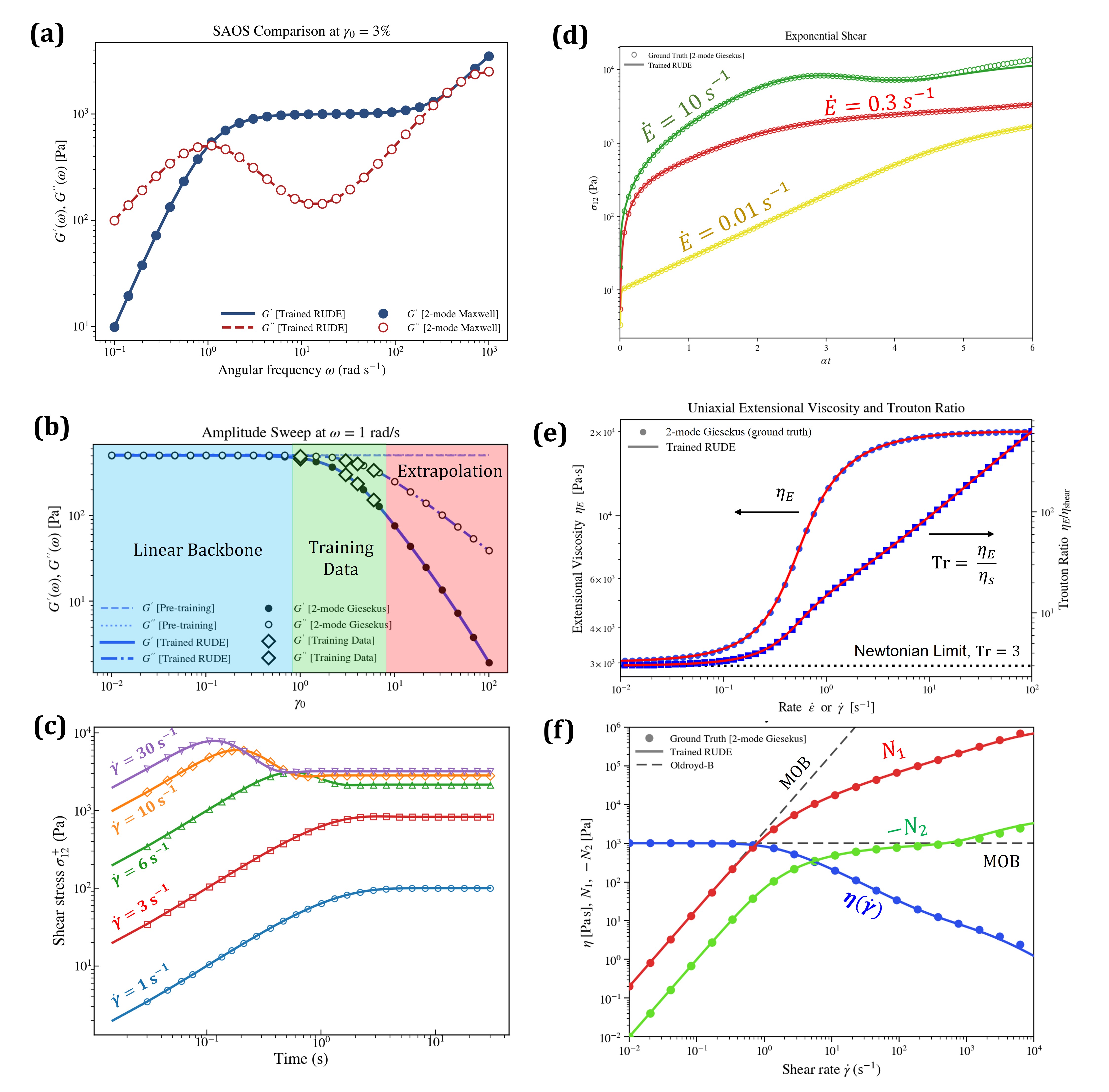}
\caption{\textbf{2-mode Giesekus with different $\alpha$ values for each mode.} The trained mmRUDEs reproduce the results of the ground truth Giesekus model (with $\alpha_1=0.1$ for the first mode and $\alpha_2=0.05$ for the second mode) across oscillatory and steady flows. In all panels, symbols represent the reference Giesekus model, solid lines show the trained mmRUDEs prediction, and dashed lines indicate the MOB backbone before training (pre-training, $\bF=\bm{0}$). (\textbf{a})~Comparison of the linear viscoelastic spectra predicted by the trained model and the ground truth, showing the storage modulus $G'(\omega)$ and loss modulus $G''(\omega)$; (\textbf{b})~Large-amplitude oscillatory shear (LAOS) at $\omega=1$~rad/s as the strain amplitude ($\gamma_0$) is varied. The green regime is the training window ($\gamma_0=1,3,4,6$; $0.995 \leq \mathrm{Wi}_{\bar\tau_\eta}\leq 5.97$); the blue regime (low amplitude) shows that the learned correction vanishes in the linear limit; the red regime (up to $\gamma_0=100$, $\mathrm{Wi}_{\bar\tau_\eta}\approx99.5$) shows accurate extrapolation for nonlinear flows. (\textbf{c})~Shear stress growth predictions and ground truth for the startup of steady shear from the initial transient to the steady-state responses exhibit good agreement. (\textbf{d})~ Extrapolation for the trained mmRUDEs on 2-mode Giesekus data to an unseen strong flow history, exponential shear. (\textbf{e})~Steady uniaxial extensional viscosity and Trouton ratio over four decades of extension rate. (\textbf{f})~Evolution of the steady first normal stress difference $N_1(\dot\gamma)$, second normal stress difference $N_2(\dot{\gamma})$ and shear viscosity $\eta(\dot\gamma)$ over six decades of steady shear rate ($\dot\gamma=10^{-2}$--$10^{4}$~s$^{-1}$,
$0.995 \times 10^{-2} \leq \mathrm{Wi}_{\bar\tau_\eta}=\bar\tau_\eta\dot\gamma \leq 0.995 \times 10^4$); }
\end{figure*}

\newpage

\appendix

\section*{SI Appendix E: Performance Evaluation: Exponential Phan-Thien-Tanner Model}

The exponential Phan-Thien-Tanner model \cite{phan-thien_new_1977} has the following form:
\begin{equation}
    \tau \overset{\nabla}{\pmb{\sigma}} + \exp \left(\frac{\varepsilon \tau}{\eta_0}\text{tr}(\pmb{\sigma}) \right) \pmb{\sigma} + \tau \frac{\xi}{2} (\pmb{\sigma} \cdot \pmb{\dot{\gamma}} + \pmb{\dot{\gamma}} \cdot \pmb{\sigma}) = \eta_0 \dot{\pmb{\gamma}}
\end{equation}
where the model nonlinearity is controlled by two parameters. The first is the parameter $\varepsilon$, which is a measure of the stress-enhanced rate of rupture events in the transient entangled network, and the second is the parameter $\xi$, which quantifies the degree of non-affine slip between polymer strands and the macroscopic strain in the fluid. The model simplifies to the quasilinear UCM model when $\varepsilon = \xi = 0$. 

The linear viscoelastic spectra is described by two modes: (1000 Pa, 1 s), (5000 Pa, 0.001 s). Both modes have the same nonlinear parameters $\varepsilon = 0.25$ and $\xi = 0.10$. The viscosity-weighted relaxation time is $\bar{\tau}_\eta = 0.995$ s. 

The mmRUDE was trained on LAOS data at an angular frequency of $\omega = 1$ rad/s, for strain amplitudes $\gamma_0 =$ 1, 3, 4, 6 and 10. The trained mmRUDE is able to reproduce the ground truth data well for the $\omega = 1$ rad/s, but the predictions become poorer at larger strain amplitudes at a higher frequency $\omega = 2 \pi$ rad/s. The predictions for the steady shear viscosity remain reliable for high strain rates, but for $N_1$ and $-N_2$ start to deviate from the ground truth. These observations are not surprising since the mmRUDE is only trained on weakly nonlinear data. The \emph{rheome} for the trained mmRUDE also shows that it does not recover exactly the expected form of the ePTT model. However, importantly, the trained mmRUDE remains stable even when subject to strongly nonlinear deformations. 

\begin{figure*}[htbp]
\centering
\includegraphics[width=0.9\textwidth]{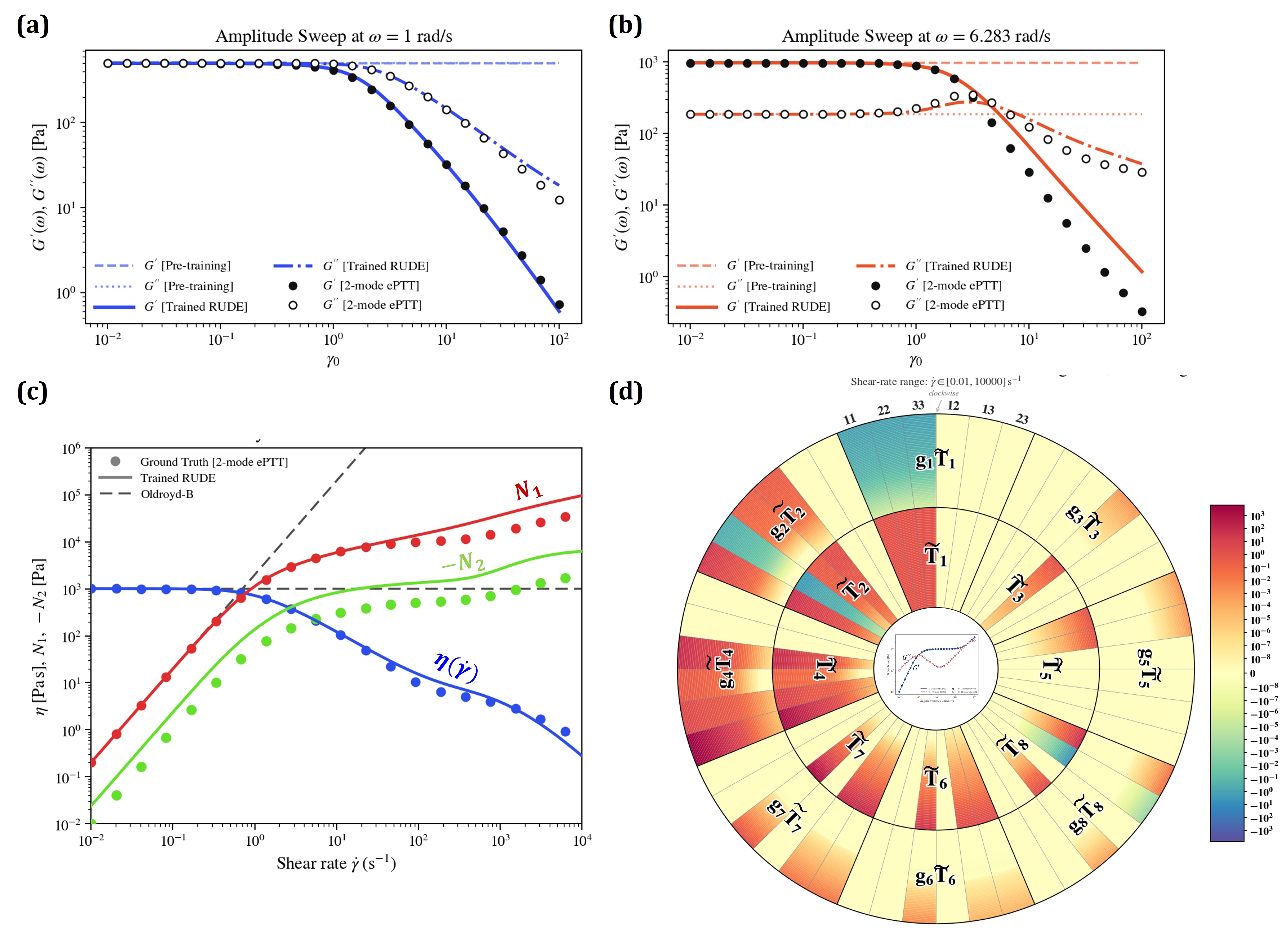}
\caption{\textbf{2-mode exponential Phan-Thien-Tanner model with the same nonlinearity for each mode.} (\textbf{a})~Large-amplitude oscillatory shear (LAOS) at $\omega=1$~rad/s as the strain amplitude ($\gamma_0$) is varied. (\textbf{a})~Large-amplitude oscillatory shear (LAOS) at $\omega=2\pi$~rad/s as the strain amplitude ($\gamma_0$) is varied. (\textbf{c})~Steady first normal stress difference $N_1(\dot\gamma)$, second normal stress difference $-N_2(\dot\gamma)$ and shear viscosity $\eta(\dot\gamma)$ over six decades of steady shear rate ($\dot\gamma=10^{-2}$--$10^{4}$~s$^{-1}$). (\textbf{d})~The \emph{rheome} in steady shear for the trained mmRUDE.}
\label{fig:ePTT}
\end{figure*}

\newpage

\section*{SI Appendix F: Commonly Used Rheological Constitutive Equations Represented in the RUDEs Framework}
\addcontentsline{toc}{section}{SI Appendix F}

Several commonly used rheological constitutive equations can be represented within the RUDEs framework. These are summarized in the table below:

\begin{table} [h]
    \centering
    \begin{tabular}{c|c|c}
        Model & General Equation & Relevant Terms in RUDEs \\ \hline
        
        Upper Convected Maxwell & $\pmb{\sigma} + \tau \overset{\nabla}{\pmb{\sigma}} = \eta_0 \pmb{\dot{\gamma}}$ & $g_n = 0$ for all $n$  \\
        
        Co-Rotational Maxwell & $\pmb{\sigma} + \tau \overset{\kern0.25em\circ}{\pmb{\sigma}} = \eta_0 \pmb{\dot{\gamma}}$ & $g_6 = \frac{1}{2} \tau$  \\

        Lower Convected Maxwell & $\pmb{\sigma} + \tau \overset{\Delta}{\pmb{\sigma}} = \eta_0 \pmb{\dot{\gamma}}$ & $g_6 = \tau$  \\
        
        Johnson-Segalman \cite{johnson_model_1977} & $\pmb{\sigma} + \tau \overset{\kern0.25em\square}{\pmb{\sigma}} = \eta_0 \pmb{\dot{\gamma}}$ & $g_6 = \frac{\xi}{2} \tau$  \\
        
        Giesekus \cite{giesekus_simple_1982} & $\pmb{\sigma} + \tau \overset{\nabla}{\pmb{\sigma}} + \frac{\alpha \tau}{\eta_0} \pmb{\sigma} \cdot \pmb{\sigma} = \eta_0 \pmb{\dot{\gamma}}$ & $g_4 = \frac{\alpha \tau}{\eta_0}$ \\[1.0em]
        
        Linear Phan-Thien-Tanner \cite{phan-thien_new_1977} & $\tau \overset{\kern0.25em\square}{\pmb{\sigma}} + (1 + \frac{\varepsilon \tau}{\eta_0}\text{tr}(\pmb{\sigma})) \pmb{\sigma} = \eta_0 \dot{\pmb{\gamma}}$ & \makecell{$g_2 = \frac{\varepsilon \tau}{\eta_0}\text{tr}(\pmb{\sigma})$, \\ $g_6 = \frac{\xi}{2} \tau$} \\[1.0em]
        
        Exponential Phan-Thien-Tanner \cite{phan-thien_new_1977} & $\tau \overset{\kern0.25em\square}{\pmb{\sigma}} + \exp \left(\frac{\varepsilon \tau}{\eta_0}\text{tr}(\pmb{\sigma})) \right) \pmb{\sigma} = \eta_0 \dot{\pmb{\gamma}}$ & \makecell{$g_2 = \exp \left(\frac{\varepsilon \tau}{\eta_0}\text{tr}(\pmb{\sigma})) \right) - 1$, \\ $g_6 = \frac{\xi}{2} \tau$} \\[1.5em]
        
        Rolie-Poly \cite{likhtman_simple_2003} & $\tau_d \overset{\nabla}{\pmb{\sigma}} + \pmb{\sigma} + \frac{2\tau_d}{\tau_R} (1 - \lambda^{-1}) (G \pmb{I} + (1 + \beta \lambda^{2\delta}) \pmb{\sigma}) = G \tau_d \dot{\pmb{\gamma}}$ & \makecell{$g_1 = \frac{2\tau_d}{\tau_R} (1 - \lambda^{-1}) G$, \\ $g_2 = \frac{2\tau_d}{\tau_R} (1 - \lambda^{-1})(1 + \beta \lambda^{2\delta})$}
    \end{tabular}
    \label{tab:placeholder}
\end{table}

Most of these models are parameterized by a zero-shear viscosity $\eta_0$ and a single relaxation time $\tau$. 

The upper convected derivative of the stress tensor is: 
\begin{align}
    \overset{\nabla}{\pmb{\sigma}} \equiv \frac{D \pmb{\sigma}}{D t} - (\nabla \pmb{v})^T \cdot \pmb{\sigma} - \pmb{\sigma} \cdot \nabla \pmb{v}
\end{align}

The lower convected derivative of the stress tensor is:
\begin{align}
    \overset{\Delta}{\pmb{\sigma}} \equiv \frac{D \pmb{\sigma}}{D t} + (\nabla \pmb{v}) \cdot \pmb{\sigma} + \pmb{\sigma} \cdot (\nabla \pmb{v})^T
\end{align}

The Gordon-Schowalter derivative is related to the upper and lower convected derivatives as shown:
\begin{equation}
    \overset{\kern0.25em\square}{\pmb{\sigma}} = \overset{\nabla}{\pmb{\sigma}} + \frac{\xi}{2} (\pmb{\sigma} \cdot \pmb{\dot{\gamma}} + \pmb{\dot{\gamma}} \cdot \pmb{\sigma})
\end{equation}
This becomes the upper convected derivative for $\xi = 0$, the co-rotational derivative for $\xi = 1$ and the lower convected derivative for $\xi = 2$.

For the Giesekus model, the nonlinearity is controlled by a mobility parameter $\alpha$. For the PTT models, the nonlinearity is controlled by a network softening parameter $\varepsilon$ and a slip parameter $\xi$. 

The Rolie-Poly model is typically expressed in terms of the polymer configuration tensor $\pmb{c}$ and the identity tensor $\pmb{I}$:
\begin{equation}
    \overset{\nabla}{\pmb{c}} = - \frac{1}{\tau_d} (\pmb{c} - \pmb{I}) - \frac{2 k_s (\lambda)}{\tau_R} \left(1 - \frac{1}{\lambda}\right) \left(\pmb{c} + \beta \lambda^{2\delta} (\pmb{c} - \pmb{I}) \right)
\end{equation}
where the tube disengagement time is $\tau_d$, and the chain stretch $\lambda = \sqrt{\text{tr}~\pmb{c} / 3}$ evolves separately with the Rouse time $\tau_R$. Two scalar model parameters $\beta$ and $\delta$ modulate the effectiveness of convective constraint release and the chain sensitivity to stretching. The tensorial stress is related to the polymer configuration tensor $\pmb{c}$ and the plateau modulus $G$:
\begin{equation}
    \pmb{\sigma} = G (\pmb{c} - \pmb{I})
\end{equation} 
An algebraic rearrangement to express the constitutive equation only in terms of the stress tensor and the strain rate tensor, and noting that $\overset{\nabla}{\pmb{I}} = - \pmb{\dot{\gamma}}$, yields the equation as presented in the table above. 

\end{document}